\RequirePackage[pagewise]{lineno}

\documentclass[aps,prd, twocolumn,superscriptaddress,amsmath,amssymb]{revtex4}

\usepackage{graphicx} 
\usepackage{epstopdf} 
\usepackage{dcolumn} 
\usepackage{xcolor} 
\usepackage{amsmath,amssymb} 
\usepackage{hyperref} 
\usepackage[utf8]{inputenc} 
\usepackage[english]{babel} 
\usepackage{blindtext} 
\usepackage{ifthen} 
\usepackage{orcidlink} 

\usepackage[11pt]{moresize}

\def\mystrut{\rule{0pt}{0.9\normalbaselineskip}}

\usepackage{siunitx}
\usepackage{subcaption}
\usepackage{turnpageetex}
\usepackage{multirow}
\usepackage{afterpage}
\usepackage{placeins}

\newboolean{articletitles}
\setboolean{articletitles}{true} 

\input{belle2-symbols}

\newcommand{\Vub}{$|V_{ub}|$\xspace}
\newcommand{\qtwo}{\ensuremath{q^2}\xspace}

\newcommand*{\thetamiss}{\ensuremath{\theta_{\mathrm{miss}}}\xspace}
\newcommand*{\cosThetaBY}{\ensuremath{\cos \theta_{\mathrm{BY}}}\xspace}
\newcommand*{\mpipi}{\ensuremath{m_{\pi\pi}}\xspace}

\newcommand*{\deltaE}{\ensuremath{\Delta E}\xspace}
\newcommand*{\Mbc}{\ensuremath{M_{\mathrm{bc}}}\xspace}

\newcommand{\Btopilnu}{\ensuremath{B^0\to\pi^- \ell^+ \nu_{\ell}}\xspace}
\newcommand{\Btorholnu}{\ensuremath{B^+\to\rho^0 \ell^+\nu_{\ell}}\xspace}
\newcommand{\totrholnu}{\ensuremath{B\to\rho \ell \nu_{\ell}}\xspace}
\newcommand{\totpilnu}{\ensuremath{B\to\pi\ell \nu_{\ell}}\xspace}
\newcommand{\Bptopilnu}{\ensuremath{B^+\to\pi^0 \ell^+ \nu_{\ell}}\xspace}
\newcommand{\Bztorholnu}{\ensuremath{B^0\to\rho^- \ell^+\nu_{\ell}}\xspace}

\newcommand{\BtoXlnu}{\ensuremath{B\to X \ell \nu_{\ell}}\xspace}

\newcommand{\BtoXclnu}{\ensuremath{B\to X_c \ell \nu_{\ell}}\xspace}
\newcommand{\BtoDlnu}{\ensuremath{B\to D \ell \nu_{\ell}}\xspace}
\newcommand{\BtoDstlnu}{\ensuremath{B\to D^* \ell \nu_{\ell}}\xspace}

\newcommand{\Btopipilnu}{\ensuremath{B\to\pi\pi\ell \nu_{\ell}}\xspace}
\newcommand{\BtoXulnu}{\ensuremath{B\to X_u \ell \nu_{\ell}}\xspace}
\newcommand{\BztoXulnu}{\ensuremath{B^0\to X_{u}^{-} \ell^+ \nu_{\ell}}\xspace}
\newcommand{\BptoXulnu}{\ensuremath{B^+\to X_{u}^{0} \ell^+ \nu_{\ell}}\xspace}
\newcommand{\Btoomlnu}{\ensuremath{B\to \omega \ell \nu_{\ell}}\xspace}
\newcommand{\Btoetalnu}{\ensuremath{B\to \eta \ell \nu_{\ell}}\xspace}
\newcommand{\Btoetaplnu}{\ensuremath{B\to \eta^{\prime} \ell \nu_{\ell}}\xspace}
\newcommand{\totetalnu}{\ensuremath{B\to \eta^{(\prime)} \ell \nu_{\ell}}\xspace}

\newcommand{\Gap}{\ensuremath{B\rightarrow D^{(*)}\eta \ell\nu_{\ell}}\xspace}

\newcommand{\BB}{\ensuremath{B\overline{B}}\xspace}
\newcommand{\qq}{\ensuremath{q\overline{q}}\xspace}

\begin{document}


\vspace{6pt}

\title{Determination of $|V_{ub}|$ from simultaneous measurements of untagged $B^0\to\pi^- \ell^+ \nu_{\ell}$ and $B^+\to\rho^0 \ell^+\nu_{\ell}$ decays}
  \author{I.~Adachi\,\orcidlink{0000-0003-2287-0173}} 
  \author{L.~Aggarwal\,\orcidlink{0000-0002-0909-7537}} 
  \author{H.~Aihara\,\orcidlink{0000-0002-1907-5964}} 
  \author{N.~Akopov\,\orcidlink{0000-0002-4425-2096}} 
  \author{A.~Aloisio\,\orcidlink{0000-0002-3883-6693}} 
  \author{N.~Althubiti\,\orcidlink{0000-0003-1513-0409}} 
  \author{N.~Anh~Ky\,\orcidlink{0000-0003-0471-197X}} 
  \author{D.~M.~Asner\,\orcidlink{0000-0002-1586-5790}} 
  \author{H.~Atmacan\,\orcidlink{0000-0003-2435-501X}} 
  \author{T.~Aushev\,\orcidlink{0000-0002-6347-7055}} 
% ...
\collaboration{The Belle II Collaboration} 

\begin{abstract}
We present a measurement of $|V_{ub}|$ from a simultaneous study of the charmless semileptonic decays $B^0\to\pi^- \ell^+ \nu_{\ell}$ and $B^+\to\rho^0 \ell^+\nu_{\ell}$, where $\ell = e, \mu$. This measurement uses a data sample of 387 million $B\overline{B}$ meson pairs recorded by the Belle~II detector at the SuperKEKB electron-positron collider between 2019 and 2022.
The two decays are reconstructed without identifying the partner $B$ mesons. We simultaneously measure the differential branching fractions  of $B^0\to\pi^- \ell^+ \nu_{\ell}$ and $B^+\to\rho^0 \ell^+\nu_{\ell}$ decays as functions of $q^2$ (momentum transfer squared).
From these, we obtain total branching fractions $\mathcal{B}(B^0\to\pi^- \ell^+ \nu_{\ell}) = (1.516 \pm 0.042 (\mathrm{stat}) \pm 0.059 (\mathrm{syst})) \times 10^{-4}$ and $\mathcal{B}(B^+\to\rho^0 \ell^+\nu_{\ell}) = (1.625 \pm 0.079 (\mathrm{stat}) \pm 0.180 (\mathrm{syst})) \times 10^{-4}$.
By fitting the measured $B^0\to\pi^- \ell^+ \nu_{\ell}$ partial branching fractions as functions of $q^2$, together with constraints on the non-perturbative hadronic contribution from lattice QCD calculations, we obtain $|V_{ub}|$ = $(3.93 \pm 0.09 \pm 0.13 \pm 0.19) \times 10^{-3}$.
Here, the first uncertainty is statistical, the second is systematic, and the third is theoretical.
\end{abstract}

\maketitle

\section{Introduction}
The Cabibbo-Kobayashi-Maskawa (CKM)~\cite{CKM2, CKM} matrix elements are related to four fundamental parameters of the Standard Model (SM) of particle physics.
The magnitude of the matrix element $V_{ub}$ can be determined by measuring the rate of \BtoXulnu decays, which is proportional to $|V_{ub}|^{2}$, where $X_{u}$ is a charmless hadronic final state and $\ell$ is a light charged lepton.
Two methods can be used to measure $|V_{ub}|$. 
In the inclusive method, no specific $X_{u}$ final state is reconstructed, and the sum of all possible final states is analyzed. 
The theoretical description involves calculation of the total semileptonic rate. 
In the exclusive method, a specific final state is reconstructed and the theoretical description takes the form of parametrizations of the low-energy strong interactions (form factors).
These two methods have complementary uncertainties introduced by their theoretical descriptions.
Determinations of \Vub from experimental data obtained by the inclusive and exclusive methods currently differ by approximately 2.5 standard deviations~\cite{HFLAV}.

The most experimentally and theoretically reliable exclusive measurements of \Vub come from \totpilnu decays. 
The current world average of \Vub from this mode, taking all of the experimental and theoretical information into consideration, is $(3.67 \pm 0.09 \pm 0.12 )\times 10^{-3}$~\cite{HFLAV}, where the first uncertainty is experimental and the second is theoretical.
A simultaneous measurement of \Btopilnu with \totrholnu decays allows for an additional measurement of \Vub while accounting for decays of one type that are reconstructed as the other. 
Therefore, we simultaneously reconstruct \Btopilnu and \Btorholnu (with charge conjugation implied throughout), focusing on decays with entirely charged final states in order to avoid backgrounds associated with neutral particles. 
We measure the differential branching fractions of these decays as a function of \qtwo, the squared momentum transferred from the \B meson to the hadron. 
We use data from the Belle~II detector located at the SuperKEKB electron-positron collider at KEK, in Japan.
The \B-meson decays are reconstructed in \epem \to \FourS \to \BB events.
We use an ``untagged'' reconstruction method in which the signal lepton and hadron candidates are selected without prior reconstruction (tagging) of the partner $B$ meson.
This leads to a high signal efficiency, but also a low purity due to the large combinatorial background from the partner $B$ meson and increased backgrounds from \epem collisions that produce light quark pair (continuum) events.

We reconstruct \Btopilnu and \Btorholnu events and make signal selections that reduce backgrounds. 
For both modes we build three-dimensional histograms of two kinematic variables and the reconstructed \qtwo. 
From simulation, we define \Btopilnu and \Btorholnu signal categories based on 13 and 10 disjoint intervals (bins) of simulated \qtwo, respectively. 
Signal templates defined in this way, which we refer to as true \qtwo binning, can overlap in reconstructed \qtwo due to resolution effects. 
We extract signal yields in all 23 signal categories with a simultaneous extended maximum-likelihood fit to both histograms. 
From these, we obtain differential branching fractions as functions of true \qtwo. 
We then combine constraints from theoretical form-factor predictions from unquenched lattice QCD (LQCD)~\cite{FLAG21} and light-cone sum rule (LCSR)~\cite{LCSR_pi, BSZ} calculations with the measured differential branching fractions to determine \Vub.

\section{Theoretical Framework}
In contrast to hadronic decays, in semileptonic \BtoXulnu decays, the leptonic and hadronic parts of the decay matrix element factorize at leading electroweak order, which allows for a straightforward determination of \Vub.
However, the description of the hadronic contribution to the matrix element has many uncertainties, since higher-order perturbative corrections and hadronization processes need to be included. 
Form factors parametrize these QCD processes in the full phase space and are constrained in specific phase-space regions by theoretical predictions based on, for example, LQCD and LCSR  calculations~\cite{JochenReview}. 
These predictions rely on the $V{-}A$ structure of the process and are usually formulated as functions of \qtwo.
For the semileptonic decay \BtoXlnu,
\begin{linenomath*}
\begin{equation}
    q^2=(p_B-p_{X})^2,
    \label{eq:qtwo}
\end{equation}
\end{linenomath*}
where $p_B$ and $p_X$ are the four-momenta of the $B$ meson and the hadron, respectively.
Here we use natural units with $c=1$ throughout.
LQCD predictions rely on first principles to predict the form factors for \totpilnu, and are only available in the high-\qtwo region.
LCSR predictions, on the other hand, are only applicable in the low-\qtwo region, but are available for \totpilnu and \totrholnu decays. 

In the theoretical description of the kinematic processes as a function of \qtwo, it is necessary to distinguish between decays into pseudoscalar and vector mesons, such as \totpilnu and \totrholnu, respectively. 
The hadronic matrix element of the decay \totpilnu can be written in terms of two form factors $f_+(\qtwo)$ and $f_0(\qtwo)$, where $f_0(\qtwo)$ is negligible for $\ell = e, \mu$. 
The differential decay rate can then be expressed as a function of $f_+(\qtwo)$ and \Vub~\cite{JochenReview}:
\begin{linenomath*}
\begin{equation}	
    \frac{\mathrm{d}\Gamma(\totpilnu)}{\mathrm{d}\qtwo \mathrm{d}\cos{\theta_{W\ell}}} = |V_{ub}|^2 \frac{G^2_F |\vec{p}_{\pi}|^3}{32\pi^3} \sin^2{\theta_{W\ell}}|f_+(\qtwo)|^2,
    \label{eq:drate_pi}  
\end{equation}
\end{linenomath*} where $G_F$ is the Fermi constant, $\vec{p}_{\pi}$ is the momentum of the pion in the $B$-meson rest frame, and $\theta_{W\ell}$ is the angle between the $W$ boson momentum in the $B$ rest frame and the lepton momentum in the $W$ rest frame.

The hadronic matrix element of \totrholnu is described by four form
factors, which can be reduced to three, $A_1(\qtwo)$, $A_2(\qtwo)$, and $V(\qtwo)$, for decays to light leptons ($\ell = e, \mu$).
The form factors can be rewritten in terms of the helicity amplitudes, $H_+(\qtwo)$, $H_-(\qtwo)$, and $H_0(\qtwo)$, of the $\rho$ meson.
The differential decay rate as a function of $H_{\pm}$, $H_0(\qtwo)$, and \Vub then becomes~\cite{JochenReview}:
\begin{linenomath*}
\begin{multline}
    \frac{\mathrm{d}\Gamma(\totrholnu)}{\mathrm{d}\qtwo \mathrm{d}\cos{\theta_{W\ell}}} = |V_{ub}|^2 \frac{G^2_F |\vec{p}_{\rho}|\qtwo}{128\pi^3 m_B^2} \\
    \times \biggr[ \sin^2{\theta_{W\ell}}|H_0(\qtwo)|^2 
     + (1-\cos{\theta_{W\ell}})^{2} \frac{|H_+(\qtwo)|^2}{2} \\
     + (1+\cos{\theta_{W\ell}})^{2} \frac{|H_-(\qtwo)|^2}{2}\biggr],
    \label{eq:drate_rho}
\end{multline} 
\end{linenomath*}
where $m_B$ is the mass of the $B$ meson, and $\vec{p}_{\rho}$ is the momentum of the $\rho$ meson in the $B$ rest frame.

Due to the relationship between the differential decay rate of \totpilnu (or \totrholnu), \Vub, and the respective form factors, \Vub can be extracted from measurements of the differential decay rates of \totpilnu (or \totrholnu) if the \qtwo shape and normalizations of the form factors are known.
The normalizations are provided by theoretical QCD calculations of the form factors.
Since these calculations are not available across the entire \qtwo range, the \qtwo dependence of the form factors is interpolated between the high and low-\qtwo regions using analyticity and unitarity arguments. 
One technique employs dispersion relations to expand in powers of the variable $z(\qtwo,q_0^2)$ defined as 
\begin{linenomath*}
\begin{equation}	
    z(\qtwo,q_0^2) = \frac{\sqrt{m^2_{+}-\qtwo}-\sqrt{m^2_{+}-q_0^2}}{\sqrt{m^2_{+}-\qtwo}+\sqrt{m^2_{+}-q_0^2}},
    \label{eq:exp_param}
\end{equation} 
\end{linenomath*}
where $m_+ = m_B + m_X$ is the sum of the masses of the $B$ meson and the hadron. 
According to Ref.~\cite{BCL} the optimal choice of the parameter $q_0^2$ is given by $q_0^2 = m_+(\sqrt{m_B} - \sqrt{m_X})^2$.

Two expansion parametrizations are considered in this work.
The first parametrization is the Bourrely-Caprini-Lellouch (BCL) parametrization~\cite{BCL}, which expands the form factor $f_+(\qtwo)$ using form-factor coefficients $b^+_k$ up to expansion order $K$ as:
\begin{linenomath*}
\begin{equation}	
    f_+(\qtwo) = P(\qtwo) \sum^{K-1}_{k=0}b^+_k \Bigr[ z^k - (-1)^{k-K}\frac{k}{K}z^K\Bigr].
    \label{eq:expansionBCL+}
\end{equation}
\end{linenomath*}
Here $P(\qtwo) = (1-\qtwo/m^2_R)^{-1}$ is called the inverse Blaschke factor, where the mass of the resonance $m_{R}$ depends on the allowed angular momentum and parity. 
The expansion of the form factor $f_0(\qtwo)$ using form-factor coefficients $b^0_k$ takes the form:
\begin{linenomath*}
\begin{equation}	
    f_0(\qtwo) = f_+(0) \Bigr[ 1 + \sum^{K-1}_{k=0}b^0_k  z^k \Bigr].
    \label{eq:expansionBCL0}
\end{equation}
\end{linenomath*}

The Bharucha-Straub-Zwicky (BSZ) parametrization~\cite{BSZ} instead is a series expansion around $\qtwo = 0$, using form-factor coefficients $b^i_k$, so that the form factors $f_i(\qtwo) \in \{A_1(\qtwo), A_2(\qtwo), V(\qtwo)\}$ take the form
\begin{linenomath*}
\begin{equation}	
    f_i(\qtwo) = P(\qtwo) \sum^{K-1}_{k=0} b^i_k \Bigr(z(\qtwo) - z(0)\Bigr)^k.
    \label{eq:expansionBSZ}
\end{equation}
\end{linenomath*}

By fitting the form-factor parametrizations given in Equations ~\ref{eq:expansionBCL+}, \ref{eq:expansionBCL0} and \ref{eq:expansionBSZ} to measured partial branching-fraction spectra, the form-factor coefficients $b^i_k$ can be extracted.
In addition, by adding theoretical input from LCSR or LQCD calculations, we can determine \Vub.

\section{Detector, data set, and simulation}
\subsection{Detector}

The Belle~II experiment~\cite{b2tdr} is located at SuperKEKB, which collides electrons and positrons at and near the $\Upsilon(4S)$ resonance~\cite{superkekb}. 
The Belle~II detector~\cite{b2tdr} has a cylindrical geometry and includes a two-layer silicon-pixel detector~(PXD) surrounded by a four-layer double-sided silicon-strip detector~(SVD)~\cite{SVD} and a 56-layer central drift chamber~(CDC).
These detectors reconstruct trajectories (tracks) of charged particles.
Only one sixth of the second layer of the PXD was installed for the data analyzed here. 
The symmetry axis of these detectors, defined as the $z$ axis, is almost coincident with the direction of the electron beam. 
Surrounding the CDC, which also provides $\mathrm{d}E/\mathrm{d}x$ energy-loss measurements and has a polar angle acceptance of 17--150$^{\circ}$, is a time-of-propagation counter~(TOP)~\cite{TOP} in the central region and an aerogel-based ring-imaging Cherenkov counter~(ARICH) in the forward region. 
These detectors provide charged-particle identification.
Surrounding the TOP and ARICH is an electromagnetic calorimeter~(ECL) based on CsI(Tl) crystals that primarily provides energy and timing measurements for photons and electrons. 
Outside of the ECL is a superconducting solenoid magnet. 
Its flux return is instrumented with resistive-plate chambers and plastic scintillator modules to detect muons, $K^0_L$ mesons, and neutrons. 
The solenoid magnet provides a 1.5~T magnetic field that is parallel to the $z$ axis.

Using $\mathrm{d}E/\mathrm{d}x$ energy-loss data from the CDC and information from the two particle-identification detectors, the ARICH and TOP, as well as data from the ECL, the SVD, and the $K^{0}_{L}$ and muon detector (KLM), charged particles of different masses are identified via particle-identification (PID) likelihood ratios. 
Each of these is a ratio of the likelihood $\mathcal{L}$ for one charged-particle hypothesis $\alpha$ to the sum of the likelihoods for all hypotheses: $\mathcal{R}_\alpha = \mathcal{L}_\alpha / (\mathcal{L}_e + \mathcal{L}_\mu + \mathcal{L}_\pi + \mathcal{L}_K + \mathcal{L}_p + \mathcal{L}_d)$, where $\alpha \in \{e, \mu, \pi, K$, proton ($p$), deuteron ($d$)$\}$.
The identification efficiencies and misidentification probabilities of the pion, kaon, electron, and muon likelihood ratios are determined separately for both charges in bins of momentum and polar angle using data samples of control modes, such as $J/\psi \to \ell^{+}\ell^{-}$ and $D^{*+}\rightarrow D^{0} [\to K^{-}\pi^{+}]\pi^{+}$.

\subsection{Experimental data}
The primary data set used in this analysis consists of ($387\pm 6$) million \FourS \to \BB events from $(364\pm 2)$~fb$^{-1}$ of electron-positron collisions collected at a center-of-mass (c.m.) energy of $\sqrt{s} = $ 10.58~GeV, corresponding to the $\Upsilon(4S)$ resonance.
We use an additional sample corresponding to $(42.6\pm 0.3)$~fb$^{-1}$ of off-resonance collision data, collected at a c.m.\ energy 60~MeV below the $\Upsilon$(4S) resonance, to describe background from continuum processes.
These include $q\overline{q}$ production \epem \to $u\bar u,\, d\bar d,\, s\bar s$, and $c\bar c$, QED processes such as \epem \to$\tau^+\tau^-$, and  two-photon processes such as \epem \to $e^+e^-\ell^+\ell^-$ and \epem \to $e^+e^-\pi^+\pi^-$.

\subsection{Simulation}
\label{sec:Sim}
We use simulated data sets referred to as Monte Carlo (MC) samples to identify efficient background-discriminating variables and to form fit templates for signal extraction. 
The MC samples for \BB background events correspond to an integrated luminosity of 3~ab$^{-1}$ and contain semileptonic and hadronic \B decays, generated using the \texttt{EvtGen}~\cite{EvtGen} software package.

We use generated signal MC samples containing 10 million events of \Btopilnu and \Btorholnu decays to obtain reconstruction efficiencies and study the key kinematic distributions.
In order to describe the remaining \BtoXulnu decays, we simulate four samples, each containing 50 million events, of resonant and nonresonant \BztoXulnu, and \BptoXulnu.
The resonant sample contains \BtoXulnu decays, where $X_{u} \in \{\pi,\rho,\omega,\eta,\eta'\}$.
We simulate the hadronization of the $X_u$ state to multiple hadrons in nonresonant \BtoXulnu events with \texttt{PYTHIA}~\cite{Pythia}. 
The second $B$ meson decays in the same way as the generic \BB background described above.

We also generate 1~ab$^{-1}$ of \qq and $\tau\tau$ continuum-event samples with \texttt{KKMC}~\cite{KKMC}, and simulate \qq fragmentation with \texttt{PYTHIA}~\cite{Pythia} and $\tau$ decays with \texttt{TAUOLA}~\cite{TAUOLA}.
Additionally, we simulate 2~ab$^{-1}$ of two-photon production with \texttt{AAFH}~\cite{AAFH}.
Final-state radiation of photons from stable charged particles is simulated with \texttt{PHOTOS}~\cite{PHOTOS} and we overlay simulated beam-induced backgrounds to all generated events~\cite{BeamBKG}. 
The propagation of particles through the detector and the resulting interactions are simulated using \texttt{Geant4}~\cite{GEANT4}.
We reconstruct and analyze simulated and experimental data with the \belletwo analysis software framework, \texttt{basf2}~\cite{basf2, basf2Repo}.

In the simulation of \BB backgrounds, we use world-average \BtoXlnu branching fractions~\cite{PDG} in combination with an isospin-symmetry assumption, following the procedure in Ref.~\cite{BFcalc} for \BtoXclnu decays. 
We assume that the remaining difference between the sum of the exclusive \BtoXclnu decay branching fractions and the measured total branching fraction, accounting for approximately 4\% of the \BtoXclnu decays, is saturated by \Gap decays, which corresponds to the procedure in Ref.~\cite{GapMethod}.
For the form factors of \BtoDlnu and \BtoDstlnu decays, we use the Boyd-Grinstein-Lebed parametrization (BGL)~\cite{BGL} with central values from Ref.~\cite{D_FF} and \cite{Dst_FF}, respectively.

For \BtoXulnu decays, the sum of the measured exclusive branching fractions is approximately 20\% of the inclusively measured branching fraction. 
We model the remaining difference with nonresonant \BtoXulnu decays.
An important component of these nonresonant \BtoXulnu decays are events in which the $X_u$ system consists of a $\pi^+\pi^-$ pair.
The partial-branching-fraction spectrum of $B^+ \to \pi^+ \pi^- \ell^+ \nu_{\ell}$ as a function of the di-pion invariant mass \mpipi has been measured in Ref.~\cite{Nonres}. 
We compare this experimental spectrum to the simulated spectrum of nonresonant $B^+ \to \pi^+ \pi^- \ell^+ \nu_{\ell}$ events. 
Assuming that nonresonant \BtoXulnu events have no structure below the $\rho^0$ mass peak, we interpolate the measured spectrum to the $\rho^0$ mass peak window by fitting a straight line with floating slope and intercept to the spectrum surrounding the peak. 
We then assign event weights to the simulated $B^+ \to \pi^+ \pi^- \ell^+ \nu_{\ell}$ events in order to recover the measured partial branching fractions as a function of \mpipi in the simulation.

We describe the remaining nonresonant \BtoXulnu decays with the De Fazio and Neubert (DFN) model~\cite{DFN}, which combines a prediction of the triple-differential rate in $X_{u}$ particle mass $m_{X}$, the $B$ rest-frame lepton energy $E_{\ell}$, and \qtwo, with a non-perturbative shape function using an exponential model.
In the simulation we use central values for the two relevant parameters provided in Ref.~\cite{DFNparam}.
The total \BtoXulnu composition is described by implementing a hybrid model~\cite{Hyb}, following closely the method in Ref.~\cite{munu}.
This approach combines the exclusive and nonresonant decay rates in bins of $m_{X}$, $E_{\ell}$, and \qtwo, in order to reproduce the inclusive rates. 

We describe \totpilnu decays using the BCL parametrization~\cite{BCL} with central values for the parameters $b^+_k$ and $b^0_k$ in Equations~\ref{eq:expansionBCL+} and ~\ref{eq:expansionBCL0}, respectively, from Ref~\cite{FLAG21}, and \totrholnu and \Btoomlnu decays using the BSZ parametrization~\cite{BSZ} with central values for $b^i_k$ in Equation~\ref{eq:expansionBSZ} from Ref.~\cite{rho_om_FF}. 
The lineshape of the $\rho$ meson is modeled following the description in Ref.~\cite{RhoFlorian} neglecting interference between the $\rho$ and the $\omega$ meson, which is included as a systematic uncertainty and described in Section~\ref{sec:sys}.
For the form-factor description of \totetalnu decays we use a LCSR calculation~\cite{eta_FF}.

\section{Event reconstruction and selection}
\subsection{Signal reconstruction and selection}

We begin signal reconstruction by identifying track candidates that pass certain quality criteria. 
The extrapolated trajectories must originate from a cylindrical region of length $3$~cm along the $z$ axis and radius $0.5$~cm in the transverse plane, centered on the $e^+e^-$ interaction point. 
Furthermore, charged particles must have transverse momenta greater than 0.05~GeV and polar angles within the CDC acceptance.
We discard events with fewer than five tracks satisfying the above criteria.

In the remaining events, we select signal-lepton candidates from among the selected tracks by requiring that their c.m.\ momenta $p^{*}_{\ell}$ are in the range $[1.0,2.85]$~GeV and $[1.4,2.85]$~GeV in the \Btopilnu and \Btorholnu modes, respectively. 
These selections significantly reduce the number of events in which the lepton originates from \BtoXclnu decays, where $X_{c}$ is a hadronic final state containing a charm quark.
We choose a higher lepton momentum threshold for \Btorholnu candidates, since, due to the different spin structure, the lepton momentum spectrum of \Btorholnu peaks at a higher momentum than that of \Btopilnu.
In order to reduce sensitivity to modeling of the detector response in the extreme forward and backward directions of the lepton polar angle $\theta_{\ell}$, we exclude events with $\cos \theta_{\ell} < -0.55$ and $\cos \theta_{\ell} > 0.85$.

We require that electron and muon candidates have PID likelihood ratios greater than 0.9.
The electron likelihood ratio combines information from the CDC, ECL, ARICH and the KLM, while the muon likelihood ratio also includes information from the TOP.
The average electron (muon) efficiency is 92 (93)\%.
The hadron misidentification rates are 0.2\% for the electron and 3.2\% for the muon selection, respectively.
The four-momenta of the electron candidates are corrected for bremsstrahlung by adding the four-momenta of photons with an ECL cluster energy below $1.0$~GeV for \Btopilnu and $0.5$~GeV for \Btorholnu found within a cone of 0.05~rad around the electron-momentum vector.

In the \Btopilnu mode we select pion candidates from the remaining tracks and require that they have a charge opposite to that of the lepton candidate.
In the \Btorholnu mode we require that the two selected pion candidates that compose the $\rho$ candidate have opposite charges and have a mass \mpipi in the range $[0.554,0.996]$~GeV. 
This selection reduces combinatorial background, but is loose enough to reduce sensitivity to modeling of the $\rho$ mass distribution.
We reduce sensitivity to the modeling of the detector response in the extreme backward region of pion polar angle $\theta_\pi$ by selecting pions with $\cos \theta_\pi > -0.65$.
All pion candidates are required to have a PID likelihood ratio greater than $0.1$.
The pion likelihood, in addition to data from the CDC, ECL, ARICH and the KLM, includes information from the SVD and the TOP.
To improve the particle-identification performance, we require that the pion candidates have at least 20 measurement points in the CDC.
The resulting average pion efficiency is 86\% with kaon and lepton misidentification rates of 7\% and 0.4\%, respectively. 

The selections described below are designed to reduce backgrounds and enhance signal purity. 
We remove candidates with kinematic properties inconsistent with the signal $B$ meson decay. 
Assuming that only a single massless particle is not included in the event reconstruction, the angle between the $B$ meson and the combination of the signal lepton and hadron candidates, denoted $Y$, is given by
\begin{linenomath*}
\begin{equation}	
\cosThetaBY = \frac{2E^*_{B}E^*_{Y}-m_{B}^{2}-m_{Y}^{2}}{2|\vec{p}^{\,*}_{B}||\vec{p}^{\,*}_{Y}|},
\label{eq:thetaBY}
\end{equation}
\end{linenomath*}
where $E^*_{Y}$, $|\vec{p}^{\,*}_{Y}|$, and $m_{Y}$ are the energy, magnitude of the three-momentum, and invariant mass of the $Y$ in the c.m.\ frame, respectively. 
The energy $E^*_{B}$ and the magnitude of the three-momentum $|\vec{p}^{\,*}_{B}|$ of the $B$ meson are calculated from the beam properties, and $m_{B}$ is the mass of the $B$ meson~\cite{PDG}.
For correctly reconstructed signal decays and assuming perfect resolution, we expect \cosThetaBY to lie between $-1$ and 1. 
However, to retain enough background events to train the classifiers discussed in Section~\ref{sec:BDT}, we choose a less restrictive requirement, $|\cosThetaBY| < 1.6$.
Additionally, we perform vertex fits~\cite{TreeFit} to the hadron and lepton candidates and require that they converge. 

\subsection{Missing momentum reconstruction}
We estimate the momentum of the signal neutrino by attributing the sum of the remaining tracks and electromagnetic energy depositions (clusters) in the event, called the rest of event (ROE), to the partner $B$. 
From energy and momentum conservation, we construct the missing four-momentum in the c.m.\ frame, 
\begin{linenomath*}
\begin{equation}	
(E^{*}_{\mathrm{miss}},\vec{p}^{\,*}_{\mathrm{miss}}) =  (E^*_{\Upsilon(4S)},\vec{p}^{\,*}_{\Upsilon(4S)})  - \left( \sum_{i} E^{*}_{i} , \sum_{i} \vec{p}^{\,*}_{i}\right),
\label{eq:p_miss}
\end{equation}
\end{linenomath*}
where $E^{*}_{i}$ and $\vec{p}^{\,*}_{i}$ correspond to the c.m.\ energy and momentum of the $i$th track or cluster in the event, respectively.
We determine $E^{*}_{i}$ using the momentum derived from the reconstructed track and select the mass hypothesis $\alpha$ with the highest value of the likelihood ratio $\mathcal{R}_\alpha$.
We attribute the missing four-momentum to the signal neutrino, with momentum $\vec{p}^{\,*}_{\nu} = \vec{p}^{\,*}_{\mathrm{miss}}$, and energy, $E^{*}_{\nu} = |\vec{p}^{\,*}_{\nu}| = |\vec{p}^{\,*}_{\mathrm{miss}}|$.
Taking the magnitude of $\vec{p}^{\,*}_{\mathrm{miss}}$, instead of $E^{*}_{\mathrm{miss}}$, to approximate the neutrino energy, leads to an improvement in resolution of 15\%.
While reconstruction losses add up linearly in the calculation of $E^{*}_{\mathrm{miss}}$, this is not the case for the vector sum calculation of $\vec{p}^{\,*}_{\mathrm{miss}}$.

Since all reconstructed tracks and clusters contribute to the resolution of the neutrino momentum estimation, obtaining an ROE as pure and complete as possible is critical.
To reduce the impact of clusters from beam-induced backgrounds, acceptance losses, or other effects, we impose quality requirements for objects to be included in the ROE. 
We only consider clusters that are within the CDC acceptance with energies in the forward, barrel, and backward directions greater than 0.060, 0.050, and 0.075~GeV, respectively.
We require that the clusters contain more than one calorimeter crystal and are detected within 200~ns of the collision time, which is approximately five times the mean timing resolution of the calorimeter.
In addition to removing background particles from the ROE, we must account for particles that escape undetected.
To reduce the impact of events with undetected particles, we require that the polar angle of the missing momentum in the laboratory frame \thetamiss is within the CDC acceptance.

\subsection{Signal extraction variables}
\label{sec:SigExtVar}

We reconstruct \qtwo{} from Equation~\ref{eq:qtwo}, and thus need to estimate the $B$ momentum vector.
One existing method, called the Diamond Frame~\cite{DiamondFrame}, takes the weighted average of four possible $\vec{p}^{\,*}_{B}$ vectors uniformly distributed in azimuthal angle on the cone defined by \cosThetaBY, weighting by the $\sin^2\theta_{B}$ distribution, which expresses the prior probability of the $B$ flight direction in \FourS decays with respect to the electron-positron beam axis.
A second method, called the ROE method~\cite{ROEmethod}, assumes the signal $B$ momentum vector to be the vector on the \cosThetaBY cone that is closest to antiparallel to the ROE momentum vector $\vec{p}^{\,*}_{\mathrm{ROE}}$. 
There is a third method~\cite{CombMethod} that combines these two by multiplying the Diamond Frame weights by $\frac{1}{2}(1-\hat{p}^{\,*}_B\cdot \hat{p}^{\,*}_{\mathrm{ROE}})$ and averaging over ten vectors uniformly distributed on the cone, where $\hat{p}^{\,*}_B$ and $\hat{p}^{\,*}_{\mathrm{ROE}}$ denote the unit vectors of $\vec{p}^{\,*}_{B}$ and $\vec{p}^{\,*}_{\mathrm{ROE}}$, respectively.
We adopt this combined method because, in simulation, it assigns reconstructed signal candidates to the correct \qtwo bin more often than other methods do, leading to a reduction in the bin migrations of up to 2\%. 
The resolutions in \qtwo decrease with increasing \qtwo and vary from 0.09--0.60~GeV$^{2}$ in the \Btopilnu mode, and from 0.16--0.84~GeV$^{2}$ in the \Btorholnu mode.

We divide $B$ candidates into 13 reconstructed \qtwo bins in the \Btopilnu mode and into 10 bins in the \Btorholnu mode. 
The lowest bin boundary is at zero, and the first 12 (9) bins have uniform bin widths of 2~GeV$^{2}$ in the \Btopilnu (\Btorholnu) mode. 
The last bins extend to the kinematic limits of 26.4~GeV$^{2}$ in the \Btopilnu mode and 20.3~GeV$^{2}$ in the \Btorholnu mode.
The following are the labels and bin edges for the \qtwo bins: $q1: q^2\in[0, 2]$, $q2: [2, 4]$, $q3: [4, 6]$, $q4: [6, 8]$, $q5: [8, 10]$, $q6: [10, 12]$, $q7: [12, 14]$, $q8: [14, 16]$, $q9: [16, 18]$, $q10: [18, 20(20.3)]$, $q11: [20, 22]$, $q12: [22, 24]$, $q13: [24, 26.4]$~GeV$^{2}$.

Two additional variables that test the kinematic consistency of a candidate with a signal $B$ decay using ROE information are the beam-constrained mass, defined as
\begin{linenomath*}
\begin{equation}	
    \Mbc = \sqrt{E^{*2}_{\mathrm{beam}} -|\vec{p}^{\,*}_{B}|^{2}} =\sqrt{\left(\frac{\sqrt{s}}{2 }\right)^{2} -|\vec{p}^{\,*}_{B}|^{2}}
    \label{eq:mbc}
\end{equation}
\end{linenomath*}
and the energy difference, defined as
\begin{linenomath*}
\begin{equation}
    \deltaE = E^{*}_{B} - E^{*}_{\mathrm{beam}} = E^{*}_{B} - \frac{\sqrt{s}}{2 },
    \label{eq:deltae}
\end{equation} 
\end{linenomath*}
where $E^{*}_{\mathrm{beam}}$, $E^{*}_{B}$ and $\vec{p}^{\,*}_{B}$ are the single-beam energy, the reconstructed $B$ energy, and the reconstructed $B$ momentum, all determined in the \FourS rest frame, respectively. 
The reconstructed $B$ energy (momentum) is given by the sum of the reconstructed energies (momenta) of the signal lepton and hadron candidates and the inferred neutrino energy (momentum) described above.
We define a fit region in \deltaE and \Mbc, corresponding to $-0.95 < \deltaE <1.25$~GeV and $5.095 < \Mbc < 5.295$~GeV.
This region is enriched in signal, but at the same time includes background-enhanced regions to allow sufficient discrimination between signal and background.

\section{Background suppression}
\subsection{Signal and background categories}
\label{subsec:categ}
As explained in Section~\ref{sec:Sim}, the simulated samples can be separated into two main categories: \BB events and continuum events. 
For the \BB events we define a subcategory that combines true signal and combinatorial signal events. 
In combinatorial signal either the signal hadron or lepton candidate is incorrectly chosen.
In addition, we define an isospin-conjugate signal category, where the lepton originates from the isospin conjugate of the signal decay. 
The branching fraction of the isospin-conjugate signal events scales with the branching fraction of true signal events under the assumption of isospin symmetry:
\begin{center}
    $\mathcal{B}(\Btopilnu) = 2 \tau_0/\tau_+ \times \mathcal{B}(\Bptopilnu)$, \\
    $\mathcal{B}(\Bztorholnu) = 2 \tau_0/\tau_+ \times \mathcal{B}(\Btorholnu)$, 
\end{center}
where $\tau_+/\tau_0$ is the $B$ lifetime ratio.

A fourth signal category, called cross-feed signal, includes events in which the lepton originates from the reconstructed signal decay mode, but was reconstructed in another signal mode, i.e.\ \totrholnu events reconstructed in the \totpilnu sample, and vice versa.
Since the number of true, combinatorial, and isospin-conjugate signal events scales with the same branching fraction as the cross-feed signal events from the other reconstructed mode, we combine the four signal categories into a total signal category.

We further split the remaining \BB events into the two largest semileptonic backgrounds, \BtoXclnu and \BtoXulnu. 
The remaining \BB events are combined into the \emph{other} \BB category, which is mainly composed of candidates with misidentified leptons, or with leptons from secondary decays.

\subsection{Background suppression using boosted decision trees}
\label{sec:BDT}
In order to further reduce the \BB and continuum backgrounds, we train boosted decision trees (BDTs) to separate signal from these two background categories using the FastBDT package~\cite{FastBDT}.
Since the background composition is different in the \Btopilnu and \Btorholnu modes and depends on the \qtwo bin, we train the classifier and optimize the selection separately for each \qtwo bin and mode.
In addition, to increase the sensitivity in the highest \qtwo bin in the \Btorholnu mode, we split the last bin into two bins during training and optimization of the selection.
Since separate classifiers are trained for the suppression of continuum and \BB backgrounds, we train a total of $2\times(13+11)=48$ BDTs. 
In each training we use equal amounts of simulated true signal and background events, corresponding to 400~fb$^{-1}$ of simulated continuum data, and 1~ab$^{-1}$ of simulated \BB data.

We use event-shape, kinematic and topological variables as BDT input variables.
The event-shape input variables includes four normalized Fox-Wolfram moments~\cite{FWM} and variables based on the thrust axis, which is the axis that maximizes the sum of the projected momenta of a collection of particles in the event~\cite{Thrust}.
Distinct thrust axes can be defined for the signal $B$ and the ROE.
Their magnitudes, the angle between the two axes $\cos\theta_{T}$, and the angle between the signal $B$ thrust axis and the beam direction serve as input variables.
Three cones with opening angles of 10$^{\circ}$, 20$^{\circ}$, and 30$^{\circ}$ centered around the signal $B$ thrust axis are defined, and the momentum flows into each of the three cones are added as input variables~\cite{CLEOCones}.

The kinematic and topological variables include the p-value from
the $\chi^{2}$ of the vertex fit to the pion and lepton candidates, and the cosine of the angle between the signal $B$ momentum vector and the vector connecting its fitted vertex to the interaction point, both in the plane parallel and perpendicular to the beam axis.
Other input variables are \cosThetaBY, \thetamiss, the number of tracks, the momentum of the ROE, and the polar angle of the lepton candidate. 
The remaining variables are the angle between the direction of the lepton in the $W$ frame and the $W$ in the $B$ frame, and for the \Btorholnu mode, the angle between the $\rho$ in the $B$ frame and the pion in the $\rho$ rest frame. 

We train each of the 48 BDTs using the twelve input variables that provide the highest discriminating power in the corresponding \qtwo bin, and ensure that none of the selected input variables are strongly correlated with \deltaE or \Mbc.
The input variable with the most discriminating power for the suppression of continuum events is $\cos\theta_{T}$, while \cosThetaBY and the $\chi^{2}$ vertex fit probability are the input variables with the most discriminating power for the suppression of \BB events. We identify the optimal selection criterion on the combination of the continuum and \BB output classifiers within each \qtwo bin by maximizing the ratio between the number of signal events and the square root of the sum of the number of signal and background events, as predicted by simulation.
In the \Btopilnu{}(\Btorholnu) mode these selections retain 55\%(48\%) of the signal, 4\% (1\%) of the \BB background, and 1\% (0.6\%) of the continuum background.

\subsection{Continuum reweighting}

Because of the small size of the off-resonance data, especially after suppressing continuum events, using it directly to construct fit templates results in large statistical fluctuations in the individual bins.
Instead, we weight the simulated continuum candidates using off-resonance data in order to use the resulting template during signal extraction.
To obtain the weights we initially compare the \qtwo shapes of simulated continuum data and off-resonance data, where we account for the difference in cross section between the on- and the off-resonance data sets. 
We observe similar normalization differences in the \Btopilnu and \Btorholnu modes, corresponding to a ratio between the number of events in off-resonance to simulated data of 1.2.
After correcting the simulated continuum data for the total normalization, differences in the \qtwo and \deltaE spectra remain.
These residual differences are removed by correcting the simulated two-dimensional \qtwo and \deltaE spectra using bin-by-bin event weights, which are the ratios between the number of off-resonance events and the number of simulated continuum events in each bin.
This approach relies on the assumption that the difference between off-resonance data and the simulated continuum sample is independent of \Mbc.
We validate this assumption and verify the reweighting procedure by obtaining reasonable p-values from $\chi^2$ tests on the distributions of \deltaE and \Mbc in the reweighted simulated and off-resonance data.
The systematic uncertainty due to the limited size of the off-resonance data sample is described in Section~\ref{sec:sys}.

\subsection{Selection summary}

At this stage of the analysis, an average of $1.08$ ($1.18$) candidates remain for each selected event in the \Btopilnu (\Btorholnu) mode.
Some of the events are also reconstructed in both modes. 
In events with multiple candidates in both or either of the modes, we randomly select one and discard the rest.
In this way we ensure that a single event cannot contribute multiple times to either or both of the \Btopilnu and \Btorholnu modes.

\begin{figure*}[ht!]
	\begin{subfigure}[c]{0.495\textwidth}
		\raggedleft
		\includegraphics[width=0.90\linewidth]{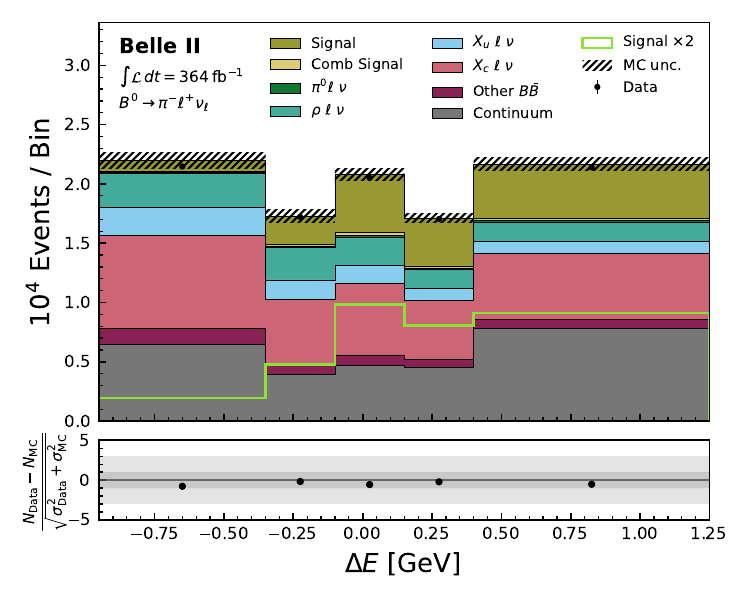}
	\end{subfigure}
	\begin{subfigure}[c]{0.495\textwidth}
		\raggedright
		\includegraphics[width=0.90\linewidth]{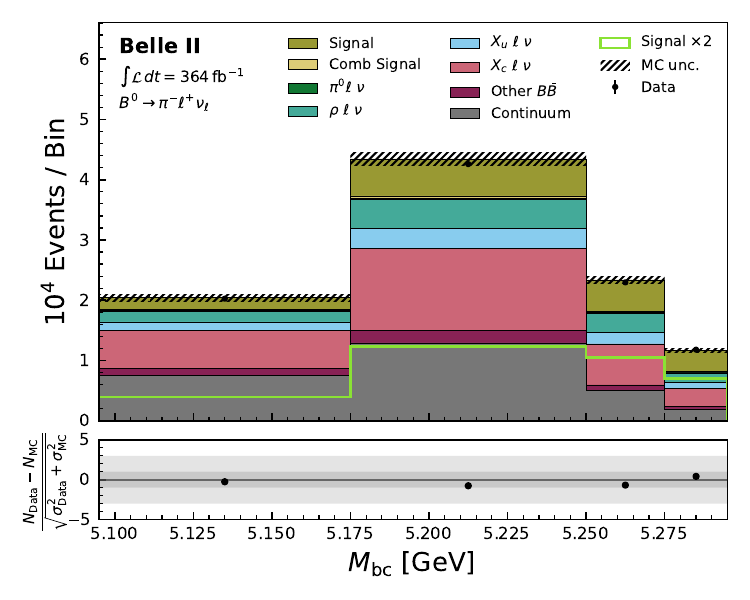}
	\end{subfigure}
        \begin{subfigure}[c]{0.495\textwidth}
		\raggedleft
		\includegraphics[width=0.90\linewidth]{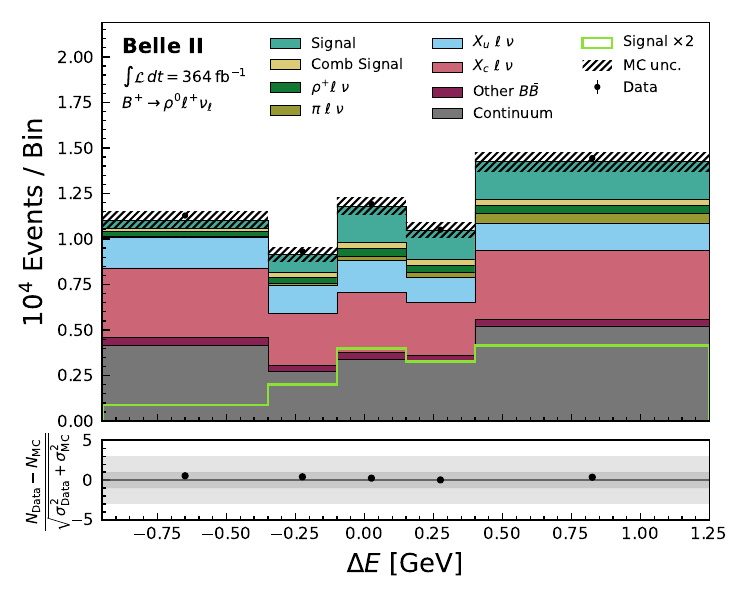}
	\end{subfigure}
	\begin{subfigure}[c]{0.495\textwidth}
		\raggedright
		\includegraphics[width=0.90\linewidth]{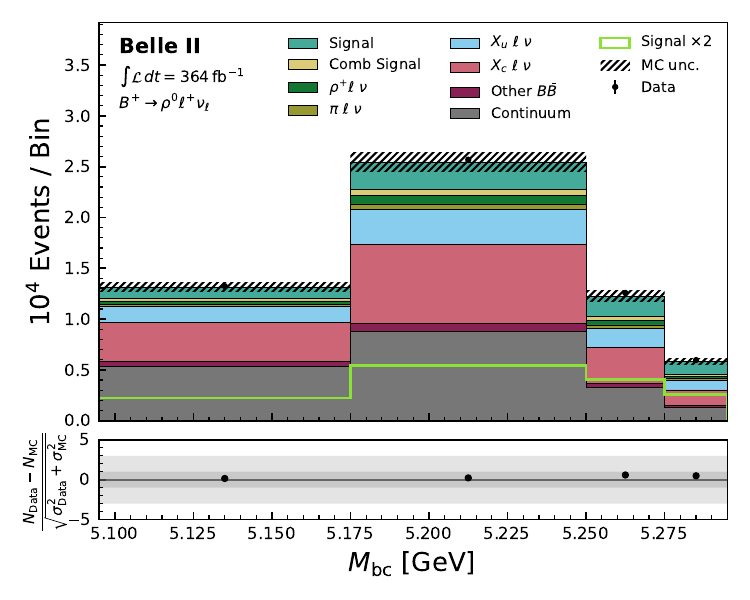}
	\end{subfigure}
	\caption{Distributions of \deltaE{} (left) and \Mbc{} (right) reconstructed in Belle~II data integrated over the \qtwo bins for \Btopilnu decays (top) and \Btorholnu decays (bottom) with expected distributions from simulation overlaid. The simulated samples are weighted according to luminosity. The hatched areas include statistical and systematic uncertainties on the simulated distributions, discussed in Section~\ref{sec:sys}. The expected signal distributions (scaled by a factor two) are also shown. The panels below the histograms show the difference between collision and simulated data divided by the combined uncertainty.}
	\label{fig:deMbc_shapes}
\end{figure*}

We divide the simulated signal events into categories based on their true \qtwo values, using the same intervals as the reconstructed \qtwo bins described in Section~\ref{sec:SigExtVar}. 
We define the signal efficiency as the fraction of signal events generated in a given true \qtwo interval that survive all selections, regardless of whether the generated events are reconstructed in the same \qtwo interval. 
In simulation, these efficiencies vary from 9\% to 19\% in the \Btopilnu mode and from 3\% to 9\% in the \Btorholnu mode. 
We call the ratio between the number of true and combinatorial signal events and the total number of signal events the signal strength.
Depending on the true \qtwo bin, the signal strength varies between 69\% and 99\% in the \Btopilnu mode and between 23\% and 57\% in the \Btorholnu mode.
The distributions of \deltaE and \Mbc integrated over the 13 (10) reconstructed \qtwo bins for \Btopilnu (\Btorholnu) decays after all selections are shown in Fig.~\ref{fig:deMbc_shapes}.

\section{Signal extraction}
\subsection{Fit method}
We extract the \Btopilnu and \Btorholnu signal yields by performing a simultaneous extended maximum likelihood fit to the binned three-dimensional distributions of \deltaE, \Mbc and reconstructed \qtwo.
We divide the \deltaE and \Mbc distributions into 5 and 4 bins, respectively.
The corresponding bin widths vary, with smaller bins being used in the signal-rich regions. 
The binning is chosen so that a clear separation of signal and background is achieved, while ensuring a sufficient number of events populate each bin. 
This results in 20 bins per reconstructed \qtwo bin, and therefore a total of 20 $\times [$13 (\Btopilnu) + 10 (\Btorholnu)$] =$ 460 bins.

The background in each signal mode is treated separately and split into four components according to the description in Section~\ref{subsec:categ}.
The fit parameters include unconstrained scale factors $b_{p}^{(\pi/\rho)}$ for each mode $(\pi/\rho) \ell \nu$ and each background component $p$.
The background templates are constructed from simulated \BB data and the reweighted continuum data.
We introduce Gaussian penalty factors to constrain the continuum yields to the scaled off-resonance yields.

In addition to the background templates, we also have one independent total signal template for each true \qtwo bin $i$ and each mode, resulting in 23 signal templates with unconstrained scale factors $s_i^{\pi}$ and $s_i^{\rho}$. 
Because these templates are based on true \qtwo categories, they naturally account for resolution effects that could cause events that are generated in one \qtwo bin to be reconstructed in another. 
The composition of the total signal component is described in Section~\ref{subsec:categ}.
In this way, the \Btopilnu background in the \Btorholnu mode is linked with the \Btopilnu signal in the \Btopilnu mode, and vice versa. 
The signal templates are constructed from simulated signal events. 
The contribution of true and combinatorial signal to the total signal templates, the signal strength, is inferred from simulation.
A summary of the 31 templates and scale factors is given in Table~\ref{tab:sca_fac}.

\begin{table}[ht!]
	\begin{center}
	    \caption{Summary of the templates and corresponding scale factors determined from the fit for the different background sources and signal samples. There is one signal scale factor for each true \qtwo bin $i$ and each signal decay, where $i \in [1,13]$ for $s_i^{\pi}$ and $i \in [1,10]$ for $s_i^{\rho}$. All fit parameters are free, with $b^{\pi}_{\mathrm{cont}}$ and $b^{\rho}_{\mathrm{cont}}$ constrained by off-resonance data.}
		\begin{tabular}{lcc}
		    \hline
		    \hline
      		\multirow{2}{*}{Component}& \multicolumn{2}{c}{Reconstructed mode} \\
			& \Btopilnu & \Btorholnu \\
			\hline
            Signal: &&  \\
			True signal &$s_i^{\pi}$& $s_i^{\rho}$ \\
			Combinatorial signal & $s_i^{\pi}$ & $s_i^{\rho}$\\
			Isospin-conjugate signal & $s_i^{\pi}$ & $s_i^{\rho}$\\
			Cross-feed &$s_i^{\rho}$ &$s_i^{\pi}$\\
                \hline \mystrut 
                Background:&& \\
                \BtoXulnu& $b_{X_u\ell\nu}^{\pi}$ & $b_{X_u\ell\nu}^{\rho}$\\
		      \BtoXclnu& $b_{X_c\ell\nu}^{\pi}$ & $b_{X_c\ell\nu}^{\rho}$\\
                Other \BB & $b_{\BB}^{\pi}$ & $b_{\BB}^{\rho}$ \\
                Continuum & $b_{\mathrm{cont}}^{\pi}$ & $b_{\mathrm{cont}}^{\rho}$\\
			\hline
			\hline
		\end{tabular}
		\label{tab:sca_fac}
	\end{center} 
\end{table}

The likelihood to be maximized is
\begin{linenomath*}
\begin{equation}	
\mathcal{L}(\vec{S},\vec{B}\,) = \prod_{l} \mathrm{Poisson}(N_{l}|\sum_{j}S_{lj}+\sum_{k}B_{lk}),
\label{eq:likelihood}
\end{equation}
\end{linenomath*}
where $N_{l}$ is the observed number of events in bin $l$, $\vec{S}$ and $\vec{B}$ are the vectors of signal and background templates, respectively, $S_{lj}$ is the number of events in bin $l$ of signal fit template $j$, and $B_{lk}$ is the number of events in bin $l$ of background fit template $k$. 

\subsection{Fit results}

\begin{figure*}[tp]
	\begin{subfigure}[c]{0.99\textwidth}
		\centering
		\includegraphics[width=0.98\linewidth]{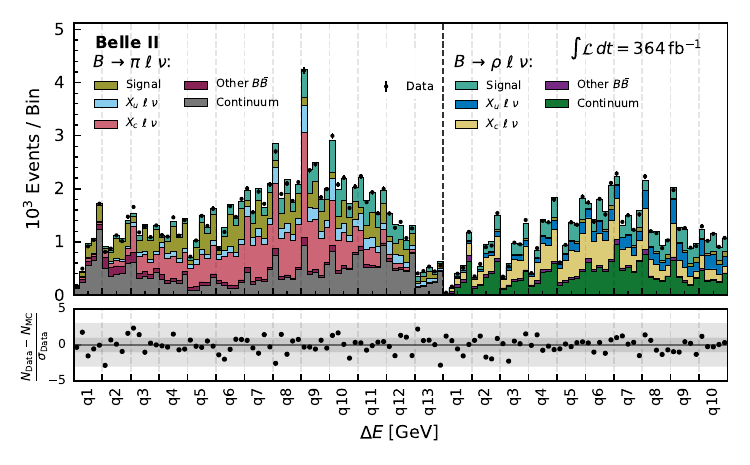}
	\end{subfigure}
	\begin{subfigure}[c]{0.99\textwidth}
		\centering
		\includegraphics[width=0.98\linewidth]{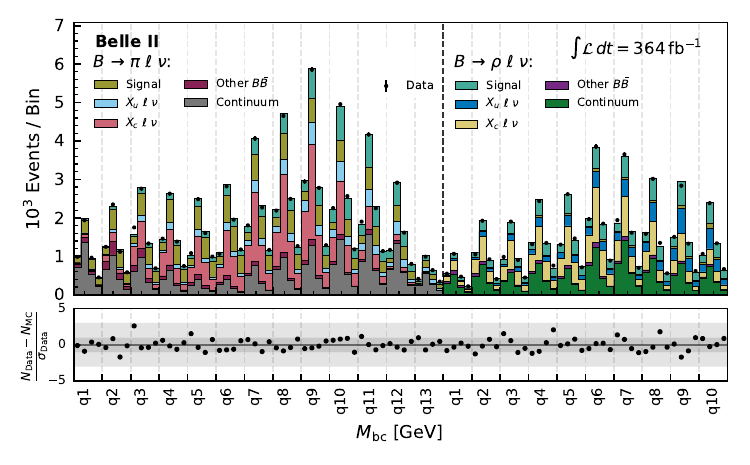}
	\end{subfigure}
	\caption{Distributions of \deltaE (top) and \Mbc (bottom) in the \qtwo bins for \Btopilnu and \Btorholnu candidates reconstructed in Belle~II data with fit projections overlaid. The difference between collision and simulated data divided by the collision data uncertainty is shown in the panels below the histograms. The boundaries of the \qtwo bins are provided in the text above.}
	\label{fig:fitprojections}
\end{figure*}

The fit projections of \deltaE and \Mbc in each \qtwo bin are shown in Fig.~\ref{fig:fitprojections} for the \Btopilnu and \Btorholnu modes.
The $\chi^2$ per degree of freedom of the fit is $468.5/429 = 1.09$. 
The correlations between the component yields are all smaller than $0.75$.
The highest observed correlations occur between the \BtoXclnu and \BB background yields in the \Btorholnu mode.
In the higher \qtwo bins, the signal scale factor becomes increasingly correlated to the \BtoXulnu scale factor.

Using the expected number of signal events from simulation, the fitted signal scale factors, and the signal strengths, we obtain the signal yields in each true \qtwo bin, corresponding to the number of true and combinatorial signal events.
The signal yields with statistical and systematic uncertainties are given in Table~\ref{tab:sig_yields}. 
The sources of systematic uncertainty and their estimation is described in Section~\ref{sec:sys}.

\begin{table}[ht!]
	\begin{center}
	    \caption{Signal yields for the \Btopilnu and \Btorholnu modes in each true \qtwo bin with statistical and systematic uncertainties. The boundaries of the \qtwo bins are given in the text above.}
		\begin{tabular}{ccc}
		    \hline
		    \hline
                \quad\quad & \multicolumn{2}{c}{Yield} \\
			\qtwo bin \quad\quad & \Btopilnu \quad\quad & \Btorholnu \\
			\hline
		      $q1\phantom{0}$ \quad\quad & \phantom{0}869 $\pm$ \phantom{0}95 $\pm$ 139 \quad\quad 
            & \phantom{0}332 $\pm$ 100 $\pm$ 118\\
			$q2\phantom{0}$ \quad\quad & 1406 $\pm$ 123 $\pm$ 172 \quad\quad & \phantom{0}651 $\pm$ 114 $\pm$ 178\\
			$q3\phantom{0}$ \quad\quad & 1426 $\pm$ 112 $\pm$ 124 \quad\quad & \phantom{0}630 $\pm$ 131 $\pm$ 131\\
			$q4\phantom{0}$ \quad\quad &1714 $\pm$ 120 $\pm$ 139
            \quad\quad & 1028 $\pm$ 147 $\pm$ 240\\
			$q5\phantom{0}$ \quad\quad & 1617 $\pm$ 120 $\pm$ 113 
            \quad\quad & 1273 $\pm$ 158 $\pm$ 236\\
			$q6\phantom{0}$ \quad\quad & 2167 $\pm$ 138 $\pm$ 151 
            \quad\quad & 1207 $\pm$ 164 $\pm$ 244\\
                $q7\phantom{0}$ \quad\quad & 1817 $\pm$ 143 $\pm$ 172 
            \quad\quad & \phantom{0}962 $\pm$ 136 $\pm$ 206 \\
                $q8\phantom{0}$ \quad\quad & 1921 $\pm$ 147 $\pm$ 181 
            \quad\quad & 1141 $\pm$ 118 $\pm$ 218\\
                $q9\phantom{0}$ \quad\quad & 1640 $\pm$ 149 $\pm$ 174 
            \quad\quad & \phantom{0}936 $\pm$ 114 $\pm$ 186\\
                $q10$ \quad\quad & 1328 $\pm$ 142 $\pm$ 156  
            \quad\quad & \phantom{0}821 $\pm$ \phantom{0}96 $\pm$ 220\\
                $q11$ \quad\quad & 1472 $\pm$ 140 $\pm$ 239 
            \quad\quad & \\
                $q12$ \quad\quad & \phantom{0}819 $\pm$ 120 $\pm$ 211 
            \quad\quad & \\
                $q13$ \quad\quad & \phantom{0}295 $\pm$ \phantom{0}66 $\pm$ 122  
            \quad\quad & \\
			\hline
			\hline
		\end{tabular}
		\label{tab:sig_yields}
	\end{center} 
\end{table}

The partial branching fraction in true \qtwo bin $i$ is calculated using the signal yield, $N_{i}$, and the corresponding signal efficiency, $\epsilon_{i}$, from 
\begin{linenomath*}
\begin{subequations}
\label{eq:deltaBF}
\begin{align}
    &\Delta \mathcal{B}_{i}(\Btopilnu) = \frac{N_{i} (1 + f_{+0})}{4\epsilon_{i}\times N_{\BB}}, \label{eq:deltaBF_pi} \\
    &\Delta \mathcal{B}_{i}(\Btorholnu) = \frac{N_{i} (1 + f_{+0})}{4 \epsilon_{i} \times N_{\BB}} \times \frac{1}{f_{+0} }, \label{eq:deltaBF_rho} 
\end{align} 
\end{subequations}
\end{linenomath*}
where $f_{+0} = \mathcal{B}(\Upsilon (4S)\rightarrow B^{+}{B^{-}})/\mathcal{B}(\Upsilon (4S)\rightarrow B^{0}\overline{B^{0}}) = 1.065 \pm 0.052$~\cite{fp0}, and $N_{\BB}$ is the number of \BB pairs.
The partial branching-fraction $\Delta \mathcal{B}$ results for \Btopilnu and \Btorholnu decays are given in Table~\ref{tab:Full_BF} and shown as functions of \qtwo in Fig.~\ref{fig:Vub}.
The entries in the total correlation matrices of the partial branching fractions of \Btopilnu and \Btorholnu are presented in Tables~\ref{tab:BFcovtot_pi} and~\ref{tab:BFcovtot_rho} in Appendix~\ref{app:cov}.
The central values of the partial branching fractions and the statistical and systematic covariance matrices combining both modes will be made available on HEPData.

\begin{table}[ht!]
	\begin{center}
	\caption{Partial branching fractions $\Delta \mathcal{B}$ ($\times 10^{4}$) in each \qtwo bin for the \Btopilnu{} and \Btorholnu{} modes. The first uncertainty is statistical and the second is systematic. The boundaries of the \qtwo bins are provided in the text above.}
		\begin{tabular}{ccc}
		    \hline
		    \hline
                \quad\quad & \multicolumn{2}{c}{$\Delta \mathcal{B}$ ($\times 10^{4}$) } \\
			$q^{2}$ bin \quad\quad & \Btopilnu \quad\quad & \Btorholnu \\
			\hline
			$q1\phantom{0}$ \quad \quad & 0.117 $\pm$ 0.013 $\pm$ 0.019 \quad \quad  &0.109 $\pm$ 0.033 $\pm$ 0.039 \\
			$q2\phantom{0}$ \quad \quad & 0.142 $\pm$ 0.013 $\pm$ 0.018 \quad \quad  &0.140 $\pm$ 0.025 $\pm$ 0.038 \\
			$q3\phantom{0}$ \quad \quad & 0.119 $\pm$ 0.009 $\pm$ 0.011 \quad \quad  &0.113 $\pm$ 0.024 $\pm$ 0.024   \\
			$q4\phantom{0}$ \quad \quad & 0.137 $\pm$ 0.010 $\pm$ 0.012 \quad \quad  &0.162 $\pm$ 0.023 $\pm$ 0.038 \\
			$q5\phantom{0}$ \quad \quad & 0.129 $\pm$ 0.010 $\pm$ 0.010 \quad \quad  &0.193 $\pm$ 0.024 $\pm$ 0.036 \\
			$q6\phantom{0}$ \quad \quad & 0.170 $\pm$ 0.011 $\pm$ 0.013 \quad \quad  & 0.183 $\pm$ 0.025 $\pm$ 0.037  \\
			$q7\phantom{0}$ \quad \quad & 0.139 $\pm$ 0.011 $\pm$ 0.014 \quad \quad  & 0.161 $\pm$ 0.023 $\pm$ 0.035 \\
			$q8\phantom{0}$ \quad \quad & 0.146 $\pm$ 0.011 $\pm$ 0.015 \quad \quad  &0.225 $\pm$ 0.023 $\pm$ 0.044 \\
			$q9\phantom{0}$ \quad \quad & 0.119 $\pm$ 0.011 $\pm$ 0.013 \quad \quad  &0.182 $\pm$ 0.022 $\pm$ 0.037   \\
			$q10$ \quad \quad & 0.096 $\pm$ 0.010 $\pm$ 0.012 \quad \quad   &0.158 $\pm$ 0.019 $\pm$ 0.043 \\
			$q11$ \quad \quad & 0.109 $\pm$ 0.010 $\pm$ 0.018 \quad \quad  & \\
			$q12$ \quad \quad &0.065 $\pm$ 0.010 $\pm$ 0.017 \quad \quad & \\
			$q13$ \quad \quad  &0.028 $\pm$ 0.006 $\pm$ 0.011   \quad \quad    &  \\
			\hline
			\hline
		\end{tabular}
		\label{tab:Full_BF}
	\end{center} 
\end{table}

The total branching fractions determined from the sums of the partial branching fractions are
\begin{center}
	$\mathcal{B}(\Btopilnu) = (1.516 \pm 0.042 \pm 0.059) \times 10^{-4}$ \\
        $\mathcal{B}(\Btorholnu) = (1.625 \pm 0.079 \pm 0.180) \times 10^{-4}$
\end{center}
where the first uncertainties are statistical and the second are systematic.
The results are consistent with world averages~\cite{PDG}, and the full experimental correlation between the two values is $-0.16$.

We perform additional fits to test the stability of the results.
We divide the data set by lepton flavor, by lepton charge, and by \thetamiss region.
We then fit separately for each subsample and check that the results agree within statistical uncertainties.

\section{Systematic uncertainties}
~\label{sec:sys}
The fractional uncertainties on the partial branching fractions in each \qtwo bin from various sources of systematic uncertainty are given in Tables~\ref{tab:sig_systematics_pi} and \ref{tab:sig_systematics_rho}.
All systematic uncertainties are evaluated using the same approach.
For each source of uncertainty, we vary the templates 1000 times by sampling from Gaussian distributions of the central values fully taking correlations into account.
For example, to evaluate the uncertainties due to the \Btoomlnu form factors, we sample 1000 alternative \BtoXulnu distributions by assuming the form-factor parameter uncertainties follow Gaussian distributions. 
We create 1000 simplified simulated data (toy) distributions by adding the resulting variations to the remaining nominal templates.
Finally, we fit the nominal templates to the toy distributions and obtain a covariance matrix of the fitted yields for each source of uncertainty using Pearson correlations~\cite{Pearson}.
Covariance matrices for the signal strengths and efficiencies are evaluated using a similar approach.
The systematic uncertainties on the partial branching fractions are evaluated by propagating the covariance matrices of the fitted yields, the signal strengths and efficiencies. 

\begin{table*}[ht!]
	\begin{center}
	\caption{Summary of fractional uncertainties in \% on the \Btopilnu partial branching fractions $\Delta \mathcal{B}$ in each \qtwo bin. The boundaries of the \qtwo bins are provided in the text above.}
    \begin{tabular}{lccccccccccccc}
        \hline
        \hline
            \multicolumn{14}{c}{\Btopilnu} \\
    	Source \quad & $q1$ \quad & $q2$ \quad & $q3$ \quad & $q4$ \quad & $q5$ \quad & $q6$ \quad & $q7$ \quad & $q8$ \quad & $q9$ \quad & $q10$ \quad & $q11$ \quad & $q12$ \quad &$q13$ \\
    	\hline
    	Detector effects \quad & \phantom{0}2.0 \quad & \phantom{0}0.9 \quad & \phantom{0}1.1 \quad & \phantom{0}1.0 \quad & \phantom{0}1.0 \quad & \phantom{0}1.1 \quad & \phantom{0}1.1 \quad & \phantom{0}1.0 \quad & \phantom{0}0.9 \quad & \phantom{0}1.2 \quad & \phantom{0}2.3 \quad & \phantom{0}4.1 \quad & \phantom{0}5.8 \\
            Beam energy \quad & \phantom{0}0.6 \quad & \phantom{0}0.8 \quad & \phantom{0}0.7 \quad & \phantom{0}0.8 \quad & \phantom{0}0.7 \quad & \phantom{0}0.6 \quad & \phantom{0}0.6 \quad & \phantom{0}0.6 \quad & \phantom{0}0.5 \quad & \phantom{0}0.5 \quad & \phantom{0}0.5 \quad & \phantom{0}0.6 \quad & \phantom{0}0.7 \\
    	Simulated sample size \quad & \phantom{0}4.7 \quad & \phantom{0}3.8 \quad & \phantom{0}3.3 \quad & \phantom{0}3.2 \quad & \phantom{0}3.2 \quad & \phantom{0}2.9 \quad & \phantom{0}3.8 \quad & \phantom{0}3.7 \quad & \phantom{0}4.0 \quad & \phantom{0}4.5 \quad & \phantom{0}5.9 \quad & \phantom{0}8.0 \quad &  13.6 \\
            BDT efficiency \quad & \phantom{0}1.3 \quad & \phantom{0}1.3 \quad & \phantom{0}1.3 \quad & \phantom{0}1.3 \quad & \phantom{0}1.3 \quad & \phantom{0}1.3 \quad & \phantom{0}1.3 \quad & \phantom{0}1.3 \quad & \phantom{0}1.3 \quad & \phantom{0}1.3 \quad & \phantom{0}1.3 \quad & \phantom{0}1.3 \quad & \phantom{0}1.3   \\
            Physics constraints \quad & \phantom{0}2.9 \quad & \phantom{0}2.9 \quad & \phantom{0}2.9 \quad & \phantom{0}2.9 \quad & \phantom{0}2.9 \quad & \phantom{0}2.9 \quad & \phantom{0}2.9 \quad & \phantom{0}2.9 \quad & \phantom{0}2.9 \quad & \phantom{0}2.9 \quad & \phantom{0}2.9 \quad & \phantom{0}2.9 \quad & \phantom{0}2.9 \\
    	\hline
            Signal model \quad & \phantom{0}0.1 \quad & \phantom{0}0.1 \quad & \phantom{0}0.2 \quad & \phantom{0}0.1 \quad & \phantom{0}0.0 \quad & \phantom{0}0.2 \quad & \phantom{0}0.2 \quad & \phantom{0}0.4 \quad & \phantom{0}0.3 \quad & \phantom{0}0.8 \quad & \phantom{0}0.9 \quad & \phantom{0}0.2 \quad & \phantom{0}4.9 \\
            $\rho$ lineshape \quad & \phantom{0}0.1 \quad & \phantom{0}0.1 \quad & \phantom{0}0.3 \quad & \phantom{0}0.3 \quad & \phantom{0}0.2 \quad & \phantom{0}0.1 \quad & \phantom{0}0.3 \quad & \phantom{0}0.1 \quad & \phantom{0}0.3 \quad & \phantom{0}0.1 \quad & \phantom{0}0.2 \quad & \phantom{0}0.2 \quad & \phantom{0}0.6 \\
            \hline
    	Nonresonant \Btopipilnu \quad & \phantom{0}0.5 \quad & \phantom{0}0.6 \quad & \phantom{0}0.4 \quad & \phantom{0}0.4 \quad & \phantom{0}0.5 \quad & \phantom{0}1.0 \quad & \phantom{0}1.2 \quad & \phantom{0}1.0 \quad & \phantom{0}0.8 \quad & \phantom{0}1.8 \quad & \phantom{0}1.2 \quad & \phantom{0}2.3 \quad & 14.3   \\
    	DFN parameters \quad & \phantom{0}0.8 \quad & \phantom{0}0.4 \quad & \phantom{0}1.5 \quad & \phantom{0}1.6 \quad & \phantom{0}1.4 \quad & \phantom{0}1.7 \quad & \phantom{0}1.2 \quad & \phantom{0}0.1 \quad & \phantom{0}0.7 \quad & \phantom{0}1.2 \quad & \phantom{0}2.9 \quad & \phantom{0}3.5 \quad & \phantom{0}3.7 \\
            \BtoXulnu model \quad & \phantom{0}0.2 \quad & \phantom{0}0.4 \quad & \phantom{0}0.3 \quad & \phantom{0}0.4 \quad & \phantom{0}0.2 \quad & \phantom{0}0.9 \quad & \phantom{0}1.1 \quad & \phantom{0}1.2 \quad & \phantom{0}1.0 \quad & \phantom{0}1.3 \quad & \phantom{0}1.6 \quad & \phantom{0}0.7 \quad & \phantom{0}8.7 \\
        \hline  
        \BtoXclnu model \quad & \phantom{0}1.4 \quad & \phantom{0}2.0 \quad & \phantom{0}1.7 \quad & \phantom{0}1.3 \quad & \phantom{0}1.3 \quad & \phantom{0}1.4 \quad & \phantom{0}1.8 \quad & \phantom{0}1.6 \quad & \phantom{0}1.3 \quad & \phantom{0}1.4 \quad & \phantom{0}1.1 \quad & \phantom{0}0.5 \quad & \phantom{0}1.7  \\
        \hline 
        Continuum  \quad & 15.1 \quad & 11.3 \quad & \phantom{0}7.6 \quad & \phantom{0}7.1 \quad & \phantom{0}5.8 \quad & \phantom{0}5.7 \quad & \phantom{0}8.1 \quad & \phantom{0}8.3 \quad & \phantom{0}9.6 \quad & 10.4 \quad & 14.5 \quad & 23.8 \quad & 34.4  \\
    	\hline
    	Total systematic \quad & 16.4 \quad & 12.6 \quad & \phantom{0}9.3 \quad & \phantom{0}8.7 \quad & \phantom{0}7.7 \quad & \phantom{0}7.7 \quad & 10.0 \quad & \phantom{0}9.9 \quad & 11.1 \quad & 12.2 \quad & 16.6 \quad & 26.0 \quad & 41.6  \\
    	Statistical \quad & 11.0 \quad & \phantom{0}8.8 \quad & \phantom{0}7.9 \quad & \phantom{0}7.0 \quad & \phantom{0}7.5 \quad & \phantom{0}6.4 \quad & \phantom{0}7.9 \quad & \phantom{0}7.7 \quad & \phantom{0}9.1 \quad  & 10.7 \quad & \phantom{0}9.6 \quad & 14.6 \quad  & 22.6  \\
    	\hline	
    	Total \quad & 19.7 \quad & 15.4 \quad & 12.2 \quad & 11.2 \quad & 10.7 \quad & 10.0 \quad & 12.7 \quad & 12.6 \quad & 14.4 \quad & 16.3 \quad & 19.1 \quad & 29.8 \quad & 47.3  \\
    	\hline
    	\hline
    \end{tabular}
	\label{tab:sig_systematics_pi}
	\end{center} 
\end{table*}

\begin{table*}[ht!]
    \begin{center}
    \caption{Summary of fractional uncertainties in \% on the \Btorholnu partial branching fractions $\Delta \mathcal{B}$ in each \qtwo bin. The boundaries of the \qtwo bins are provided in the text above.}
    \begin{tabular}{lcccccccccc}
        \hline
        \hline
            \multicolumn{11}{c}{\Btorholnu} \\
    	Source \quad & $q1$ \quad & $q2$ \quad & $q3$ \quad & $q4$ \quad & $q5$ \quad & $q6$ \quad &$q7$ \quad & $q8$ \quad & $q9$ \quad & $q10$   \\
    	\hline
    	Detector effects \quad &  \phantom{0}2.8 \quad &  \phantom{0}2.0 \quad &  \phantom{0}1.6 \quad &  \phantom{0}1.1 \quad &  \phantom{0}1.7 \quad &  \phantom{0}1.9 \quad &  \phantom{0}2.4 \quad &  \phantom{0}1.4 \quad &  \phantom{0}1.4 \quad &  \phantom{0}1.6   \\
            Beam energy \quad &  \phantom{0}2.1 \quad &  \phantom{0}1.9 \quad &  \phantom{0}1.9 \quad &  \phantom{0}1.5 \quad &  \phantom{0}1.3 \quad &  \phantom{0}1.1 \quad &  \phantom{0}1.0 \quad &  \phantom{0}0.9 \quad &  \phantom{0}0.8 \quad &  \phantom{0}0.5\\
    	Simulated sample size \quad &  14.1 \quad &  \phantom{0}7.8 \quad &  \phantom{0}7.4 \quad &  \phantom{0}6.3 \quad &  \phantom{0}6.3 \quad &  \phantom{0}5.2 \quad &  \phantom{0}6.4 \quad &  \phantom{0}5.6 \quad &  \phantom{0}6.2 \quad &  \phantom{0}7.3   \\
            BDT efficiency  \quad &  \phantom{0}1.6 \quad &  \phantom{0}1.6 \quad &  \phantom{0}1.6 \quad &  \phantom{0}1.6 \quad &  \phantom{0}1.6 \quad &  \phantom{0}1.6 \quad &  \phantom{0}1.6 \quad &  \phantom{0}1.6 \quad &  \phantom{0}1.6 \quad &  \phantom{0}1.6 \\
            Physics constraints \quad &  \phantom{0}2.8 \quad &  \phantom{0}2.8 \quad &  \phantom{0}2.8 \quad &  \phantom{0}2.8 \quad &  \phantom{0}2.8 \quad &  \phantom{0}2.8 \quad &  \phantom{0}2.8 \quad &  \phantom{0}2.8 \quad &  \phantom{0}2.8 \quad &  \phantom{0}2.8  \\
    	\hline
            Signal model \quad  &  \phantom{0}0.7 \quad &  \phantom{0}0.2 \quad &  \phantom{0}0.2 \quad &  \phantom{0}0.2 \quad &  \phantom{0}0.3 \quad &  \phantom{0}0.4 \quad &  \phantom{0}0.5 \quad &  \phantom{0}0.3 \quad &  \phantom{0}1.8 \quad &  \phantom{0}2.4  \\
            $\rho$ lineshape \quad &  \phantom{0}1.7 \quad &  \phantom{0}1.6 \quad &  \phantom{0}2.0 \quad &  \phantom{0}1.0 \quad &  \phantom{0}1.9 \quad &  \phantom{0}1.8 \quad &  \phantom{0}1.4 \quad &  \phantom{0}0.9 \quad &  \phantom{0}1.6 \quad &  \phantom{0}1.7 \\
            \hline
    	Nonresonant \Btopipilnu  \quad &  \phantom{0}5.6 \quad &  \phantom{0}6.3 \quad &  \phantom{0}6.7 \quad &  \phantom{0}8.6 \quad &  \phantom{0}9.3 \quad &  10.7 \quad &  10.1 \quad &  \phantom{0}7.0 \quad &  \phantom{0}7.8 \quad &  11.8 \\
    	DFN parameters \quad &  \phantom{0}3.6 \quad  &  \phantom{0}5.5 \quad  &  \phantom{0}4.1 \quad &  \phantom{0}3.5 \quad  &  \phantom{0}1.1 \quad &  \phantom{0}1.2 \quad &  \phantom{0}2.7 \quad &  \phantom{0}1.7 \quad &  \phantom{0}1.9 \quad &  \phantom{0}2.3 \\
            \BtoXulnu model \quad & \phantom{0}1.7 \quad & \phantom{0}3.0 \quad & \phantom{0}3.8 \quad & \phantom{0}5.0 \quad & \phantom{0}5.8 \quad & \phantom{0}6.1 \quad & \phantom{0}6.3 \quad & \phantom{0}1.9 \quad & \phantom{0}7.2 \quad & 12.4 \\
            \hline  
            \BtoXclnu model \quad & \phantom{0}1.8 \quad & \phantom{0}1.9 \quad & \phantom{0}1.7 \quad & \phantom{0}1.1 \quad & \phantom{0}1.4 \quad & \phantom{0}1.7 \quad & \phantom{0}0.9 \quad & \phantom{0}0.9 \quad & \phantom{0}1.9 \quad & \phantom{0}2.6 \\
            \hline 
            Continuum  \quad & 31.5 \quad & 24.3 \quad & 17.0 \quad & 19.6 \quad & 13.2 \quad & 14.8 \quad & 16.0 \quad & 16.6 \quad & 15.2 \quad & 18.7 \\
    	\hline
    	Total systematic \quad & 35.6 \quad & 27.5 \quad & 21.0 \quad & 23.5 \quad & 18.8 \quad & 20.5 \quad & 21.6 \quad & 19.4 \quad & 20.2 \quad & 27.0   \\
    	Statistical \quad & 30.0 \quad & 17.5 \quad & 20.8 \quad & 14.4 \quad & 12.4 \quad & 13.6 \quad & 14.1 \quad & 10.4 \quad & 12.2 \quad & 11.8 \\
    	\hline	
    	Total \quad & 46.6 \quad & 32.6 \quad & 29.6 \quad & 27.6 \quad & 22.6 \quad & 24.6 \quad & 25.8 \quad & 22.0 \quad & 23.6 \quad & 29.5  \\
    	\hline
    	\hline
    \end{tabular}
    \label{tab:sig_systematics_rho}
    \end{center} 
\end{table*}

\subsection{Detector and beam-energy effects}
The detector uncertainties include uncertainties arising from the tracking efficiency and the corrections to the lepton- and pion-identification efficiencies. 
All of these were estimated from studies of independent data control samples.

In addition, we observe a dependence of the reconstructed \qtwo resolution on the c.m.\ energy.
Since the c.m.\ energy in the simulated sample differs from the mean c.m.\ energy in data, we account for the effect on the shape of the signal template.
We investigate the effect using a control mode, in which we fully reconstruct $B^+\to J/\psi [\to \mu^+\mu^-] K^+$ events.
By removing one of the muons, we obtain events that are similar to signal decays with a single missing neutrino.

We find a 4\% difference in \qtwo resolution between measured and simulated data in this control mode.
We scale the resolutions in each \qtwo bin obtained from the simulated sample, and using the true \qtwo values, in combination with Gaussian smearing according to the new resolutions, produce 1000 pseudo-reconstructed \qtwo distributions.
By combining these with the remaining unaffected templates we obtain 1000 varied toy distributions, which are then fit using the nominal templates.

\subsection{Simulated sample size}
The effect of having limited samples of simulated data is considered.
The largest uncertainty contribution comes from the limited size of the simulated continuum sample.
In addition to shape variations due to the number of events within each bin, we also account for migration effects in the true \qtwo distribution, which results in signal-template migrations.
To estimate this uncertainty, we sample, with replacement, true \qtwo values from the total signal component 1000 times, split these into the true-\qtwo templates, then fit these templates to the sum of the nominal templates.

\subsection{BDT efficiency}

We estimate an uncertainty to account for possible disagreements between the signal efficiencies in experimental and simulated data of the selection on the 48 BDT output classifiers.
For each BDT output classifier selection, we use the $B^+\to J/\psi [\to \mu^+\mu^-] K^+$ control mode discussed above to determine the ratio between the efficiency in experimental and simulated data. 
The ratios are in agreement with unity within uncertainties. 

To account for these uncertainties we separately evaluate the standard deviations of the ratios for each type of BDT and mode and assign these as uncertainties on the signal efficiencies.
We obtain uncertainties of 1.1\% and 0.6\% on the efficiencies of the selection on the continuum suppression BDTs in the \Btopilnu and \Btorholnu modes, respectively. 
For the \BB suppression BDTs the uncertainties on the efficiencies are 0.7\% and 1.5\% for the \Btopilnu and \Btorholnu modes, respectively. 

\subsection{Physics constraints}
We consider additional systematic uncertainties from the number of \BB pairs $N_{\BB}$ and the branching fraction ratio of \FourS \to \BB, $f_{+0}$. 
These affect the calculation of the partial branching fractions from the yields. The uncertainty on $N_{\BB}$ results in a relative uncertainty of 1.4\%, while the uncertainty on $f_{+0}$ contributes uncertainties of 2.5\% and 2.4\% to the uncertainties on the partial branching fractions for \Btopilnu and \Btorholnu, respectively.

In addition, we assign an uncertainty to the assumption of isospin symmetry.
The assumption discussed in Section~\ref{subsec:categ} relies on the $B$ lifetime ratio $\tau_+/\tau_0 = 1.076 \pm 0.004$~\cite{PDG}.
To estimate the effect on the signal yields, we vary the relative fraction of neutral and charged $B$ modes in the signal templates by sampling Gaussian variations of the fractions within the relative uncertainty of 0.4\%. 

\subsection{Signal model and $\rho$ lineshape}
There are three ways in which the signal form-factor and branching-fraction uncertainties may affect our results.
The first is the residual signal form-factor model dependence of the signal templates.
Since the signal is extracted in multiple bins of true \qtwo, and therefore the fit is allowed to determine the \qtwo spectrum, this contribution is small but not negligible.
The second effect comes from signal form-factor and branching-fraction uncertainties on the composition of the background \BtoXulnu template, which propagate through the fit to uncertainties on the signal yields. 

The third effect accounts for bin migrations between the true \qtwo bins due to signal form-factor uncertainties.
These are reflected in uncertainties on the signal strengths.
The assigned uncertainties are smaller than the ones originating from the dependence of the background \BtoXulnu template on the signal model, but they are larger than the ones due to the residual signal-template dependence. 
The combination of these uncertainties is included in the signal model category in Tables~\ref{tab:sig_systematics_pi} and~\ref{tab:sig_systematics_rho}.

There is an additional uncertainty source concerning the modeling of the \Btorholnu signal related to the lineshape of the $\rho$ meson. 
We account for possible $\rho$-$\omega$ interference resulting in a change in the $\rho$ lineshape. 
In Ref.~\cite{RhoFlorian} an amplitude fit incorporating the interference term is performed to the di-pion invariant mass spectrum measured in Ref.~\cite{Nonres} and 1$\sigma$ variations of the diagonalized fit-parameter uncertainties are obtained.
We vary the lineshape of all true $\rho$ mesons for each variation, repeat the fit, and take the largest change in fit results from the model without $\rho$-$\omega$ interference as the systematic uncertainty.

\subsection{\BtoXulnu background}

\subsubsection{Nonresonant \Btopipilnu component}
One contribution to the \BtoXulnu background uncertainties is related to the treatment of the nonresonant \Btopipilnu component, which was described in Section~\ref{sec:Sim}.
In addition to the measured partial branching-fraction spectrum, Ref.~\cite{Nonres} also provides the corresponding covariance matrix.
We therefore vary the partial branching fractions according to this covariance matrix to determine the uncertainty on our signal yields. 
Overall, this results in the largest contribution to the uncertainties originating from the \BtoXulnu background.

\subsubsection{DFN parameters}
A second uncertainty component comes from the uncertainty in the DFN shape function parameters that define the shape of the nonresonant \BtoXulnu background. 
We follow the procedure of Ref.~\cite{HFLAV} to evaluate the uncertainties on the relevant parameters provided in Ref.~\cite{DFNparam}.
We then generate simulated samples of nonresonant \BtoXulnu events based on these varied parameters and reweight our nominal samples to match the varied samples.
We then estimate the uncertainties by sampling 1000 toy distributions in agreement with Gaussian variations of these parameters.
We also compare to a different nonresonant \BtoXulnu model~\cite{BLNP} but refrain from adding any additional uncertainties due to this, since the current uncertainties already cover any difference introduced by the change in model.

\subsubsection{\BtoXulnu model}
Finally, we evaluate the uncertainties due to the \Btoomlnu, \Btoetalnu, and \Btoetaplnu form factors, and obtain uncertainties listed in the \BtoXulnu category.
This category also includes the effects of uncertainties of the exclusive and inclusive \BtoXulnu branching fractions, except for the \totpilnu and \totrholnu branching fractions.

\subsection{\BtoXclnu model}
The \BtoXclnu model category in Tables~\ref{tab:sig_systematics_pi} and~\ref{tab:sig_systematics_rho} includes the effects of the uncertainties of the \BtoDlnu and \BtoDstlnu form-factor parameters, and the exclusive and inclusive \BtoXclnu branching fractions.
For the \Btopilnu mode this contribution to the total systematic uncertainty is larger than that of the \BtoXulnu model at low $\qtwo$, but smaller at high \qtwo.
It is subdominant over the entire \qtwo range for the \Btorholnu mode.

\subsection{Continuum reweighting}

The limited off-resonance sample size affects the continuum weights obtained during the reweighting procedure, since the weights are calculated using the number of off-resonance events within each bin. 
To account for the uncertainties of these numbers, we produce a set of 1000 continuum weights by recalculating the weights for each bin using off-resonance event numbers drawn from the corresponding Poisson distributions.
We then proceed with the usual procedure for determining systematic uncertainties described above.
The resulting uncertainties dominate both in the \Btopilnu and \Btorholnu mode, and are especially large in those \qtwo bins where the continuum background component is largest.

 \begin{figure*}[ht!]
    \begin{subfigure}[c]{0.329\textwidth}
		\centering
		\includegraphics[width=0.99\linewidth]{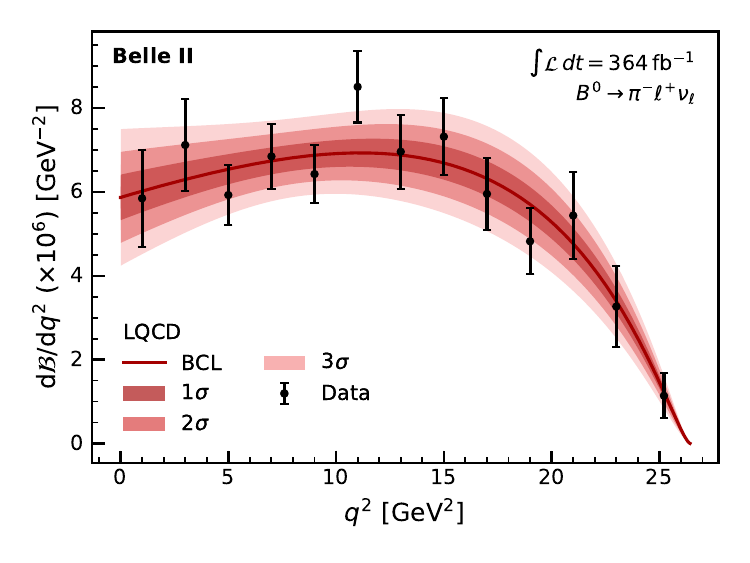}
		\subcaption{}
	\end{subfigure}
	\begin{subfigure}[c]{0.329\textwidth}
		\centering
		\includegraphics[width=0.99\linewidth]{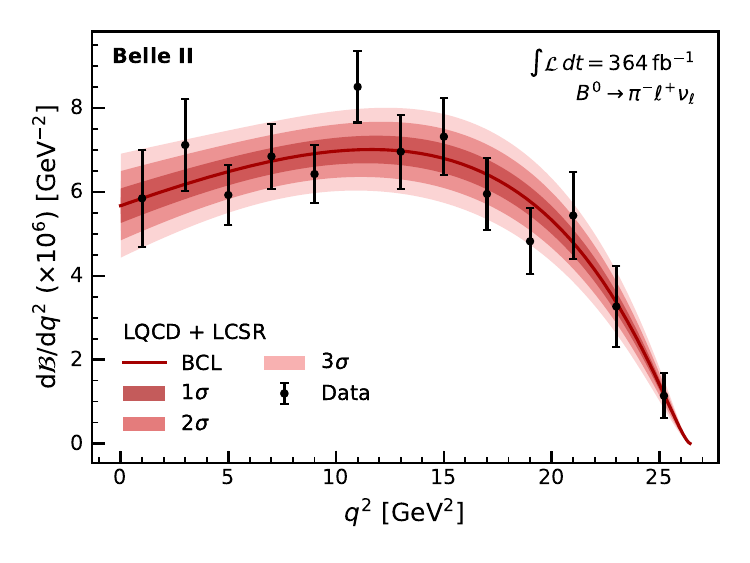}
		\subcaption{}
	\end{subfigure}
	\begin{subfigure}[c]{0.329\textwidth}
		\centering
		\includegraphics[width=0.99\linewidth]{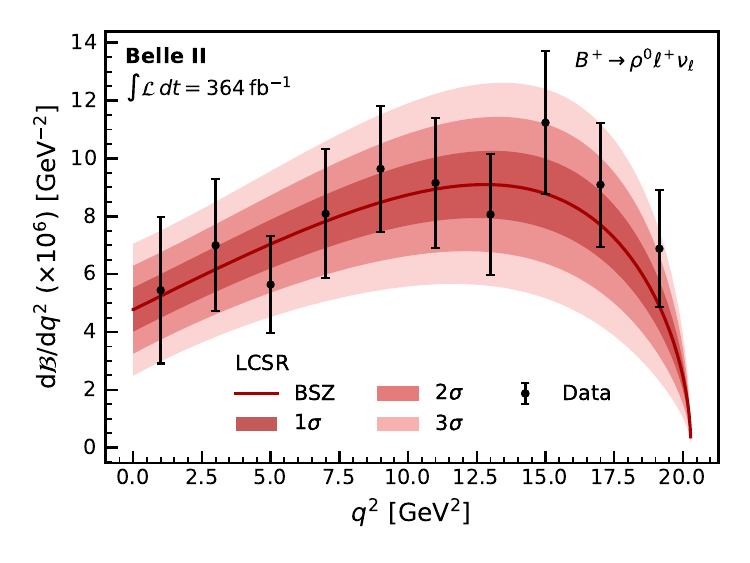}
		\subcaption{}
	\end{subfigure}
	\caption{Measured partial branching fractions as a function of \qtwo for \Btopilnu{} (a,b) and \Btorholnu (c). The fitted differential rates are shown together with the one, two, and three standard-deviation uncertainty bands for fits using constraints on the form factors from (a) LQCD, (b) LQCD and LCSR, and (c) LCSR predictions.}
    \label{fig:Vub}
\end{figure*}

\section{\Vub determination}

We extract \Vub separately from \Btopilnu and \Btorholnu using $\chi^{2}$ fits to the measured \qtwo spectra.
The $\chi^{2}$ is defined as
\begin{linenomath*}
\begin{align}
    \chi^{2} = \sum_{i,j = 1}^{N}(\Delta \mathcal{B}_{i}- \Delta \Gamma_{i}\tau) &C^{-1}_{ij} (\Delta \mathcal{B}_{j} - \Delta \Gamma_{j}\tau) \nonumber\\
 & + \sum_m \chi^{2}_{\mathrm{Theory},m},
\label{eq:minim}
\end{align} 
\end{linenomath*}
where $C^{-1}_{ij}$ is the inverse total covariance matrix of the measured partial branching fractions $\Delta \mathcal{B}_{i}$ in bin $i$, and $N$ is the number of \qtwo bins.
The quantities $\Delta\Gamma_{i}$ contain the predictions for the differential decay rates in bin $i$, $\tau$ is the $B$ lifetime, and $\chi^{2}_{\mathrm{Theory},m}$ incorporates the constraints from theory calculation $m$. 
The predictions for the differential decay rates provided in Equations~\ref{eq:drate_pi} and \ref{eq:drate_rho} include the form factors and \Vub. 

For \Btopilnu we include LQCD constraints on three $f_+(\qtwo)$ BCL form-factor coefficients $b^+_k$ and two $f_0(\qtwo)$ BCL form-factor coefficients $b^0_k$ given in Equations~\ref{eq:expansionBCL+} and \ref{eq:expansionBCL0}, respectively, as nuisance parameters.
These nuisance parameters constrain the shape and normalization of the relevant form factors entering the differential decay rate and allow for a determination of \Vub. 
In the evaluation of the inverse Blaschke factors for the expansion of $f_+(\qtwo)$ in Equation~\ref{eq:expansionBCL+}, $m_R$ takes the value of 5.325~GeV~\cite{BCL}. 
The $\chi^{2}_{\mathrm{LQCD}}$ term takes the form:
\begin{linenomath*}
\begin{align}
    \chi^{2}_{\mathrm{LQCD}} = \sum_{k,l = 1}^{5}(b_{k}-b^{\mathrm{LQCD}}_{k} ) &C^{-1}_{\mathrm{LQCD},kl} (b_{l}-b^{\mathrm{LQCD}}_{l}),
\label{eq:LQCD}
\end{align} 
\end{linenomath*}
where the constraints on the form-factor coefficients $b^{\mathrm{LQCD}}_k$ and the corresponding inverse covariance matrix $C^{-1}_{\mathrm{LQCD},kl}$ are taken from the latest version of the FLAG~21 review (February 2023)~\cite{FLAG21}, and combine results from the FNAL/MILC~\cite{FermilabMILC}, RBC/UKQCD~\cite{RBC_UKQCD}, and JLQCD~\cite{JLQCD} collaborations.

In addition to the LQCD constraints, we may also add LCSR constraints from Ref.~\cite{LCSR_pi}, which determine $f_+(\qtwo)$ and $f_0(\qtwo)$ at five points in \qtwo.
In this case the additional $\chi^{2}_{\mathrm{LCSR}}$ term takes the form:
\begin{linenomath*}
\begin{align}
    \chi^{2}_{\mathrm{LCSR}} = \sum_{k,l = 1}^{10}(f_{k}-f^{\mathrm{LCSR}}_{k} ) &C^{-1}_{\mathrm{LCSR},kl} (f_{l}-f^{\mathrm{LCSR}}_{l}).
\label{eq:LQCD}
\end{align} 
\end{linenomath*}
Here we implement direct constraints on the form factors $f_k$ ($f_+(\qtwo)$ and $f_0(\qtwo)$) from LCSR predictions $f^{\mathrm{LCSR}}_{k}$, taking the corresponding inverse covariance matrix $C^{-1}_{\mathrm{LCSR},kl}$ into account.

The result for \Vub from the \Btopilnu mode using only the LQCD constraints is:
\begin{center}
    $|V_{ub}|_{\totpilnu} = (3.93 \pm 0.09 \pm 0.13 \pm 0.19 )\times 10^{-3}$,
\end{center}
where for all quoted \Vub results the first uncertainty is statistical, the second is systematic and the third is theoretical.
Upon adding the LCSR constraints, the result for \Vub from \Btopilnu becomes:
\begin{center}
    $|V_{ub}|_{\totpilnu} = (3.73 \pm 0.07 \pm 0.07 \pm 0.16) \times 10^{-3}$.
\end{center}
The measured central values of \Vub and the BCL form-factor coefficients from the fits to the \Btopilnu spectrum are provided in Table~\ref{tab:Vub_params_pi}. 
The full correlation matrices corresponding to these values are provided in Tables~\ref{tab:LQCD_pi} and \ref{tab:LCSR_pi} in Appendix~\ref{app:cov}.

\begin{table}[ht!]
    \begin{center}
    \caption{Measured central values of \Vub and the BCL form-factor coefficients with total uncertainties from the fits to the \Btopilnu spectrum.}
    \begin{tabular}{cccc}
        \hline
        \hline
            \multicolumn{4}{c}{\Btopilnu} \\
            & \quad\quad & LQCD \quad\quad & LQCD + LCSR \\
    	\hline
    	\multicolumn{2}{c}{\Vub ($10^{-3}$)} \quad\quad& \phantom{$-$}3.93 $\pm$ 0.25 \quad\quad & \phantom{$-$}3.73 $\pm$ 0.19 \\
            \hline
            \multirow{3}{*}{$f_+(\qtwo)$} & $b_0^+$ \quad\quad & \phantom{$-$}0.42 $\pm$ 0.02  \quad\quad &  \phantom{$-$}0.45 $\pm$ 0.02 \\
    	& $b_1^+$ \quad\quad& $-$0.52 $\pm$ 0.05 \quad\quad & $-$0.52 $\pm$ 0.05  \\ 
            & $b_2^+$ \quad\quad& $-$0.81 $\pm$ 0.21 \quad\quad & $-$1.02 $\pm$ 0.18 \\
    	\hline
            \multirow{2}{*}{$f_0(\qtwo)$} & $b_0^0$ \quad\quad & \phantom{$-$}0.02 $\pm$ 0.25 \quad\quad & \phantom{$-$}0.59 $\pm$ 0.02 \\
             & $b_1^0$ \quad\quad & $-$1.43 $\pm$ 0.08 \quad\quad & $-$1.39 $\pm$ 0.07 \\
            \hline
            \multicolumn{2}{c}{$\chi^2$/ndf} \quad\quad & 8.39/7 \quad\quad & 8.36/7 \\ 
    	\hline
    	\hline
    \end{tabular}
    \label{tab:Vub_params_pi}
    \end{center} 
\end{table}

For \Btorholnu we include LCSR constraints on six BSZ coefficients $b^i_k$ given in Equation~\ref{eq:expansionBSZ} from Ref.~\cite{BSZ}. 
These correspond to constraints on two coefficients each for $A_1(\qtwo)$, $A_2(\qtwo)$, and $V(\qtwo)$.
The $\chi^{2}_{\mathrm{LCSR}}$ term for the fit to the measured \Btorholnu \qtwo spectrum takes the form:
\begin{linenomath*}
\begin{align}
    \chi^{2}_{\mathrm{LCSR}} = \sum_{k,l = 1}^{6}(b_{k}-b^{\mathrm{LCSR}}_{k} ) &C^{-1}_{\mathrm{LCSR},kl} (b_{l}-b^{\mathrm{LCSR}}_{l}),
\label{eq:LQCD}
\end{align} 
\end{linenomath*} 
where $b^{\mathrm{LCSR}}_k$ are the constraints on the coefficients and $C^{-1}_{\mathrm{LCSR},kl}$ is the corresponding inverse covariance matrix predicted by LCSR calculations. 
In the evaluation of the inverse Blaschke factors for the expansion of $A_1(\qtwo)$ and $A_2(\qtwo)$ in Equation~\ref{eq:expansionBSZ}, $m_R$ takes the value of 5.724~GeV, while it is 5.325~GeV for the expansion of $V(\qtwo)$~\cite{BSZ}. 
The \Vub result obtained from \Btorholnu using LCSR constraints is:
\begin{center}
    $|V_{ub}|_{\totrholnu} = (3.19 \pm 0.12 \pm 0.17 \pm 0.26) \times 10^{-3}$.
\end{center}
The measured central values of \Vub and the BSZ form-factor coefficients from the fit to the \Btorholnu spectrum are provided in Table~\ref{tab:Vub_params_rho}.
The full correlation matrix corresponding to these values is provided in Table~\ref{tab:LCSR_rho} in Appendix~\ref{app:cov}.
Fig.~\ref{fig:Vub} shows the measured and fitted differential rates of \Btopilnu and \Btorholnu, as well as the one, two, and three standard-deviation uncertainty bands from the fits. 

\begin{table}[ht!]
    \begin{center}
    \caption{Measured central values of \Vub and the BSZ form-factor coefficients with total uncertainties from the fit to the \Btorholnu spectrum.}
    \begin{tabular}{ccc}
        \hline
        \hline
            \multicolumn{3}{c}{\Btorholnu} \\
    	& \quad\quad & LCSR \\
    	\hline
    	\multicolumn{2}{c}{\Vub ($10^{-3}$)} \quad\quad& \phantom{$-$}3.19 $\pm$ 0.33  \\
            \hline
            \multirow{2}{*}{$A_1(\qtwo)$} & $b_0^{A_1}$ \quad\quad & \phantom{$-$}0.27 $\pm$ 0.03     \\
    	& $b_1^{A_1}$ \quad\quad & \phantom{$-$}0.34 $\pm$ 0.13   \\ 
            \hline
            \multirow{2}{*}{$A_2(\qtwo)$} & $b_0^{A_2}$ \quad\quad & \phantom{$-$}0.29 $\pm$ 0.03    \\
            & $b_1^{A_2}$ \quad\quad & \phantom{$-$}0.66 $\pm$ 0.17 \\ 
            \hline
            \multirow{2}{*}{$V(\qtwo)$} & $b_0^V$ \quad\quad & \phantom{$-$}0.33 $\pm$ 0.03  \\
            &$b_1^V$ \quad\quad & $-$0.93 $\pm$ 0.17  \\
            \hline 
            \multicolumn{2}{c}{$\chi^2$/ndf} \quad\quad & 3.85/3 \\ 
    	\hline
    	\hline
    \end{tabular}
    \label{tab:Vub_params_rho}
    \end{center} 
\end{table}

The \Vub results obtained from \Btopilnu are consistent with previous exclusive measurements~\cite{HFLAV}. 
The result obtained from \Btorholnu is lower, but consistent with previous \Vub determinations from \totrholnu decays~\cite{rho_om_FF}.
The $\chi^2$ per degree of freedom for the fits vary from 1.19 to 1.28, and are provided in Tables~\ref{tab:Vub_params_pi} and \ref{tab:Vub_params_rho} for \Btopilnu and \Btorholnu, respectively.
The extracted central values of \Vub and the coefficients, with the corresponding full covariance matrices, for the fits to the \Btopilnu and \Btorholnu spectra will be provided on HEPData.
We confirm the stability of the \Vub results by repeating the fits using different \qtwo cut-off values. 
The results are presented in Fig.~\ref{fig:Vub_stab} in Appendix~\ref{app:Vub}.

The fractional uncertainties on the \Vub results from various sources of systematic uncertainty are shown in Table~\ref{tab:Vub_systematics}.
For both \Btopilnu and \Btorholnu the largest contribution to the systematic uncertainty comes from the limited off-resonance data sample. 
In addition, for \Btorholnu the systematic uncertainty from nonresonant \Btopipilnu is significant.

\begin{table}[ht!]
    \begin{center}
    \caption{Summary of fractional uncertainties in \% on the extracted \Vub values.}
    \begin{tabular}{lccc}
        \hline
        \hline
            \quad & \multicolumn{2}{c}{\Btopilnu} & \multicolumn{1}{c}{\Btorholnu} \\
    	\quad & \multirow{2}{*}{LQCD} \quad\quad & LQCD  \quad & \multirow{2}{*}{LCSR} \\
            \quad & \quad\quad & + LCSR  \quad & \\
    	\hline
    	Detector effects \quad & 0.64 \quad\quad & 0.24 \quad & \phantom{0}0.44  \\
            Beam energy \quad & 0.05 \quad\quad &  0.03 \quad & \phantom{0}0.09   \\
    	Simulated sample size \quad &1.51 \quad\quad & 0.78 \quad & \phantom{0}1.41   \\ 
            BDT efficiency \quad & 0.31 \quad\quad & 0.21 \quad & \phantom{0}0.28 \\
            Physics constraints \quad & 0.61 \quad\quad & 0.43 \quad & \phantom{0}0.88\\
    	\hline
            Signal model \quad  &0.38 \quad\quad & 0.13 \quad & \phantom{0}0.41 \\ 
            $\rho$ lineshape \quad &0.26 \quad\quad & 0.21 \quad & \phantom{0}0.13 \\ 
            \hline
    	Nonres. \Btopipilnu \quad &0.43 \quad\quad & 0.11 \quad & \phantom{0}1.97  \\
    	DFN parameters \quad & 0.64 \quad\quad & 0.32 \quad & \phantom{0}0.88\\ 
            \BtoXulnu model \quad & 0.61 \quad\quad & 0.40 \quad & \phantom{0}1.56\\ 
            \hline  
            \BtoXclnu model \quad & 0.51 \quad\quad & 0.43 \quad & \phantom{0}0.50  \\ 
            \hline 
            Continuum \quad & 2.39 \quad\quad & 1.37 \quad & \phantom{0}4.91  \\ 
    	\hline
    	Total systematic \quad & 3.26 \quad\quad &1.91 \quad & \phantom{0}5.33  \\
    	Statistical \quad & 2.31 \quad\quad  & 1.82 \quad & \phantom{0}3.76 \\
            Theory \quad & 4.83 \quad\quad & 4.29 \quad & \phantom{0}8.15 \\
    	\hline	
    	Total \quad & 6.40 \quad\quad & 5.13 \quad &10.34  \\
    	\hline
    	\hline
    \end{tabular}
    \label{tab:Vub_systematics}
    \end{center} 
\end{table}

\section{Summary}
We extract partial branching fractions of \Btopilnu and \Btorholnu decays reconstructed in a 364~fb$^{-1}$ electron-positron collision data sample collected by the Belle~II experiment.
The branching-fraction of \Btopilnu is found to be $(1.516 \pm 0.042(\mathrm{stat}) \pm 0.059(\mathrm{syst})) \times 10^{-4}$, and the branching fraction of \Btorholnu is found to be $(1.625 \pm 0.079(\mathrm{stat}) \pm 0.180(\mathrm{syst})) \times 10^{-4}$.
These results are consistent with the current world averages~\cite{PDG}.

We extract values of the CKM matrix-element magnitude \Vub from \Btopilnu decays using LQCD constraints provided by the FLAG 21 review (updated February 2023)~\cite{FLAG21}.
We obtain $(3.93 \pm 0.09 (\mathrm{stat}) \pm 0.13 (\mathrm{syst}) \pm 0.19 (\mathrm{theo})) \times 10^{-3}$.
Using additional constraints from LCSR~\cite{LCSR_pi} this result becomes $(3.73 \pm 0.07 (\mathrm{stat}) \pm 0.07 (\mathrm{syst}) \pm 0.16 (\mathrm{theo})) \times 10^{-3}$.
The \Vub result obtained from the measurement of \Btorholnu with constraints from LCSR~\cite{BSZ} is $(3.19 \pm 0.12 (\mathrm{stat})\pm 0.18 (\mathrm{syst}) \pm 0.26 (\mathrm{theo})) \times 10^{-3}$. 

Currently our results are limited by the size of the off-resonance data set and the description of the nonresonant \BtoXulnu background. 
The first uncertainty could be reduced by improvements in the simulation of continuum backgrounds.

This work, based on data collected using the Belle~II detector, which was built and commissioned prior to March 2019,
was supported by
Higher Education and Science Committee of the Republic of Armenia Grant No.~23LCG-1C011;
Australian Research Council and Research Grants
No.~DP200101792, 
No.~DP210101900, 
No.~DP210102831, 
No.~DE220100462, 
No.~LE210100098, 
and
No.~LE230100085; 
Austrian Federal Ministry of Education, Science and Research,
Austrian Science Fund
No.~P~34529,
No.~J~4731,
No.~J~4625,
and
No.~M~3153,
and
Horizon 2020 ERC Starting Grant No.~947006 ``InterLeptons'';
Natural Sciences and Engineering Research Council of Canada, Compute Canada and CANARIE;
National Key R\&D Program of China under Contract No.~2022YFA1601903,
National Natural Science Foundation of China and Research Grants
No.~11575017,
No.~11761141009,
No.~11705209,
No.~11975076,
No.~12135005,
No.~12150004,
No.~12161141008,
and
No.~12175041,
and Shandong Provincial Natural Science Foundation Project~ZR2022JQ02;
the Czech Science Foundation Grant No.~22-18469S 
and
Charles University Grant Agency project No.~246122;
European Research Council, Seventh Framework PIEF-GA-2013-622527,
Horizon 2020 ERC-Advanced Grants No.~267104 and No.~884719,
Horizon 2020 ERC-Consolidator Grant No.~819127,
Horizon 2020 Marie Sklodowska-Curie Grant Agreement No.~700525 ``NIOBE''
and
No.~101026516,
and
Horizon 2020 Marie Sklodowska-Curie RISE project JENNIFER2 Grant Agreement No.~822070 (European grants);
L'Institut National de Physique Nucl\'{e}aire et de Physique des Particules (IN2P3) du CNRS
and
L'Agence Nationale de la Recherche (ANR) under grant ANR-21-CE31-0009 (France);
BMBF, DFG, HGF, MPG, and AvH Foundation (Germany);
Department of Atomic Energy under Project Identification No.~RTI 4002,
Department of Science and Technology,
and
UPES SEED funding programs
No.~UPES/R\&D-SEED-INFRA/17052023/01 and
No.~UPES/R\&D-SOE/20062022/06 (India);
Israel Science Foundation Grant No.~2476/17,
U.S.-Israel Binational Science Foundation Grant No.~2016113, and
Israel Ministry of Science Grant No.~3-16543;
Istituto Nazionale di Fisica Nucleare and the Research Grants BELLE2;
Japan Society for the Promotion of Science, Grant-in-Aid for Scientific Research Grants
No.~16H03968,
No.~16H03993,
No.~16H06492,
No.~16K05323,
No.~17H01133,
No.~17H05405,
No.~18K03621,
No.~18H03710,
No.~18H05226,
No.~19H00682, 
No.~20H05850,
No.~20H05858,
No.~22H00144,
No.~22K14056,
No.~22K21347,
No.~23H05433,
No.~26220706,
and
No.~26400255,
and
the Ministry of Education, Culture, Sports, Science, and Technology (MEXT) of Japan;  
National Research Foundation (NRF) of Korea Grants
No.~2016R1\-D1A1B\-02012900,
No.~2018R1\-A2B\-3003643,
No.~2018R1\-A6A1A\-06024970,
No.~2019R1\-I1A3A\-01058933,
No.~2021R1\-A6A1A\-03043957,
No.~2021R1\-F1A\-1060423,
No.~2021R1\-F1A\-1064008,
No.~2022R1\-A2C\-1003993,
and
No.~RS-2022-00197659,
Radiation Science Research Institute,
Foreign Large-Size Research Facility Application Supporting project,
the Global Science Experimental Data Hub Center of the Korea Institute of Science and Technology Information
and
KREONET/GLORIAD;
Universiti Malaya RU grant, Akademi Sains Malaysia, and Ministry of Education Malaysia;
Frontiers of Science Program Contracts
No.~FOINS-296,
No.~CB-221329,
No.~CB-236394,
No.~CB-254409,
and
No.~CB-180023, and SEP-CINVESTAV Research Grant No.~237 (Mexico);
the Polish Ministry of Science and Higher Education and the National Science Center;
the Ministry of Science and Higher Education of the Russian Federation
and
the HSE University Basic Research Program, Moscow;
University of Tabuk Research Grants
No.~S-0256-1438 and No.~S-0280-1439 (Saudi Arabia);
Slovenian Research Agency and Research Grants
No.~J1-9124
and
No.~P1-0135;
Agencia Estatal de Investigacion, Spain
Grant No.~RYC2020-029875-I
and
Generalitat Valenciana, Spain
Grant No.~CIDEGENT/2018/020;
The Knut and Alice Wallenberg Foundation (Sweden), Contracts No.~2021.0174 and No.~2021.0299;
National Science and Technology Council,
and
Ministry of Education (Taiwan);
Thailand Center of Excellence in Physics;
TUBITAK ULAKBIM (Turkey);
National Research Foundation of Ukraine, Project No.~2020.02/0257,
and
Ministry of Education and Science of Ukraine;
the U.S. National Science Foundation and Research Grants
No.~PHY-1913789 
and
No.~PHY-2111604, 
and the U.S. Department of Energy and Research Awards
No.~DE-AC06-76RLO1830, 
No.~DE-SC0007983, 
No.~DE-SC0009824, 
No.~DE-SC0009973, 
No.~DE-SC0010007, 
No.~DE-SC0010073, 
No.~DE-SC0010118, 
No.~DE-SC0010504, 
No.~DE-SC0011784, 
No.~DE-SC0012704, 
No.~DE-SC0019230, 
No.~DE-SC0021274, 
No.~DE-SC0021616, 
No.~DE-SC0022350, 
No.~DE-SC0023470; 
and
the Vietnam Academy of Science and Technology (VAST) under Grants
No.~NVCC.05.12/22-23
and
No.~DL0000.02/24-25.

These acknowledgements are not to be interpreted as an endorsement of any statement made
by any of our institutes, funding agencies, governments, or their representatives.

We thank the SuperKEKB team for delivering high-luminosity collisions;
the KEK cryogenics group for the efficient operation of the detector solenoid magnet and IBBelle on site;
the KEK Computer Research Center for on-site computing support; the NII for SINET6 network support;
and the raw-data centers hosted by BNL, DESY, GridKa, IN2P3, INFN, 
and the University of Victoria.

\bibliographystyle{belle2} 
\bibliography{references}

\appendix
\section{Correlation matrices}
\label{app:cov}
Tables~\ref{tab:BFcovtot_pi} and~\ref{tab:BFcovtot_rho} show the full experimental covariance matrices for the measurements of the \Btopilnu and \Btorholnu partial branching fractions $\Delta \mathcal{B}$, respectively.
In addition, Tables~\ref{tab:LQCD_pi} and~\ref{tab:LCSR_pi} contain the full correlation matrices of the measurements of \Vub and the form-factor coefficients from \Btopilnu using constraints from LQCD or from LQCD in combination with LCSR.
Table~\ref{tab:LCSR_rho} gives the full correlation matrix of the \Vub and form-factor coefficient measurement from \Btorholnu using LCSR constraints. 

\begin{table*}[th!]
	\begin{center}
	\caption{Total correlation matrix of the partial branching fractions $\Delta \mathcal{B}_{i}$} for \Btopilnu{}.
		\begin{tabular}{cccccccccccccc}
		    \hline
		    \hline
			\qtwo bin& $q1$ & $q2$ & $q3$ & $q4$ & $q5$& $q6$ & $q7$ & $q8$ & $q9$ & $q10$ & $q11$& $q12$ &$q13$ \\
			\hline
                $q1$ &  \phantom{$-$}1.000 &&&&&&&&&&&&\\
                $q2$ &  \phantom{$-$}0.021  &  \phantom{$-$}1.000 &&&&&&&&&&&  \\
                $q3$ &  \phantom{$-$}0.105 & $-$0.193 &  \phantom{$-$}1.000 &&&&&&&&&&\\
                $q4$ & $-$0.018 &   \phantom{$-$}0.019 & $-$0.139 & \phantom{$-$}1.000 &&&&&&&&&\\
                $q5$& $-$0.031 &$-$0.052 &   \phantom{$-$}0.202 & $-$0.053 & \phantom{$-$}1.000   &&&&&&&&\\ 
                $q6$ & \phantom{$-$}0.065 &$-$0.058 & \phantom{$-$}0.034  & \phantom{$-$}0.097 & \phantom{$-$}0.004 & \phantom{$-$}1.000  &&&&&&&\\
                $q7$ &$-$0.097 &  $-$0.160 & \phantom{$-$}0.069  &\phantom{$-$}0.226 & \phantom{$-$}0.223 & \phantom{$-$}0.090  & \phantom{$-$}1.000   &&&&&&\\
                $q8$  &$-$0.067 &$-$0.097 & \phantom{$-$}0.026 & \phantom{$-$}0.026 & \phantom{$-$}0.194 &  \phantom{$-$}0.255  & \phantom{$-$}0.213 & \phantom{$-$}1.000  &&&&&\\
                $q9$ & \phantom{$-$}0.088 & \phantom{$-$}0.035& $-$0.019& $-$0.027&  \phantom{$-$}0.053 & \phantom{$-$}0.170 &  \phantom{$-$}0.108 & \phantom{$-$}0.110 & \phantom{$-$}1.000 &&&&  \\
                $q10$ &\phantom{$-$}0.007 &$-$0.007 &  \phantom{$-$}0.001 &$-$0.053 & \phantom{$-$}0.067 &  \phantom{$-$}0.100 & \phantom{$-$}0.050 & \phantom{$-$}0.058 & \phantom{$-$}0.196 & \phantom{$-$}1.000 &&&\\
                $q11$ & \phantom{$-$}0.075 & \phantom{$-$}0.001  &   \phantom{$-$}0.059 &$-$0.005& \phantom{$-$}0.021  &\phantom{$-$}0.056 &  \phantom{$-$}0.028 &$-$0.035&  \phantom{$-$}0.148 & \phantom{$-$}0.236 &\phantom{$-$}1.000 && \\
                $q12$& \phantom{$-$}0.050  &\phantom{$-$}0.080 & \phantom{$-$}0.014  &\phantom{$-$}0.004 &$-$0.035 &$-$0.044  &$-$0.038 &$-$0.101&  \phantom{$-$}0.074  & \phantom{$-$}0.187 & \phantom{$-$}0.297 & \phantom{$-$}1.000 &\\
                $q13$ & \phantom{$-$}0.030& $-$0.053&  \phantom{$-$}0.115 &  \phantom{$-$}0.024 & \phantom{$-$}0.041 & $-$0.048 &$-$0.011 &$-$0.078 &$-$0.092 &$-$0.129 &$-$0.212 &$-$0.355 & \phantom{$-$}1.000  \\
			\hline
			\hline
		\end{tabular}
		\label{tab:BFcovtot_pi}
	\end{center} 
\end{table*}

\begin{table*}[ht!]
	\begin{center}
	\caption{Total correlation matrix of the partial branching fractions $\Delta \mathcal{B}_{i}$ for \Btorholnu{}.}
		\begin{tabular}{ccccccccccc}
		    \hline
		    \hline
			\qtwo bin& $q1$ & $q2$ & $q3$ & $q4$ & $q5$& $q6$ & $q7$ & $q8$ & $q9$ & $q10$  \\
			\hline
                $q1$ & \phantom{$-$}1.000 &&&&&&&&&\\
                $q2$ & $-$0.340 & \phantom{$-$}1.000  &&&&&&&&\\
                $q3$ &\phantom{$-$}0.146 &$-$0.322  & \phantom{$-$}1.000  &&&&&&& \\
                $q4$ &\phantom{$-$}0.023  &\phantom{$-$}0.241 &$-$0.241 & \phantom{$-$}1.000  &&&&&&\\  
                $q5$& $-$0.052  &\phantom{$-$}0.131 & \phantom{$-$}0.275& $-$0.060&  \phantom{$-$}1.000   &&&&& \\
                $q6$& \phantom{$-$}0.017 & \phantom{$-$}0.139 & \phantom{$-$}0.183 & \phantom{$-$}0.464 & \phantom{$-$}0.148  &\phantom{$-$}1.000 &&&&\\
                $q7$ & $-$0.021 &  \phantom{$-$}0.197 & \phantom{$-$}0.068 & \phantom{$-$}0.184 & \phantom{$-$}0.428& \phantom{$-$}0.030  & \phantom{$-$}1.000 &&& \\
                $q8$ & \phantom{$-$}0.149 &  \phantom{$-$}0.018 & \phantom{$-$}0.054 & \phantom{$-$}0.216 & \phantom{$-$}0.205 & \phantom{$-$}0.311 & $-$0.063 & \phantom{$-$}1.000 && \\
                $q9$ &\phantom{$-$}0.095 & \phantom{$-$}0.101 & \phantom{$-$}0.050& \phantom{$-$}0.115&   \phantom{$-$}0.136 & \phantom{$-$}0.156&  \phantom{$-$}0.235 & $-$0.005 & \phantom{$-$}1.000 &\\
                $q10$ & \phantom{$-$}0.004 &\phantom{$-$}0.187 & $-$0.083  &\phantom{$-$}0.153 & \phantom{$-$}0.151  &\phantom{$-$}0.133&  \phantom{$-$}0.188&  \phantom{$-$}0.341 &  \phantom{$-$}0.241 & \phantom{$-$}1.000   \\
			\hline
			\hline
		\end{tabular}
		\label{tab:BFcovtot_rho}
	\end{center} 
\end{table*}

\begin{table*}[ht!]
	\begin{center}
	\caption{Full correlation matrix of \Vub and the BCL form-factor coefficients from the fit to the \Btopilnu spectrum with LQCD constraints.}
		\begin{tabular}{ccccccc}
		    \hline
		    \hline
                &  \Vub  &   $b_0^+$  &  $b_1^+$   & $b_2^+$  &  $b_0^0$   & $b_1^0$ \\
                \hline
                \Vub &  \phantom{$-$}1.000 &&  &  & & \\
                $b_0^+$ & $-$0.806 &  \phantom{$-$}1.000 &&  &  & \\
                $b_1^+$&  $-$0.053  &$-$0.273 & \phantom{$-$}1.000 &&& \\
                $b_2^+$ &  \phantom{$-$}0.062 & $-$0.319 &$-$0.338 & \phantom{$-$}1.000 && \\
                $b_0^0$ & $-$0.315 &  \phantom{$-$}0.409 &$-$0.073 &$-$0.204 & \phantom{$-$}1.000 & \\
                $b_1^0$ &  $-$0.142 & $-$0.048 &  \phantom{$-$}0.150 &\phantom{$-$}0.258 &$-$0.775 & \phantom{$-$}1.000 \\
			\hline
			\hline
		\end{tabular}
		\label{tab:LQCD_pi}
	\end{center} 
\end{table*}

\begin{table*}[ht!]
	\begin{center}
	\caption{Full correlation matrix of \Vub and the BCL form-factor coefficients from the fit to the \Btopilnu spectrum with LQCD and LCSR constraints.}
		\begin{tabular}{ccccccc}
		    \hline
		    \hline
                &  \Vub  &   $b_0^+$  &  $b_1^+$   & $b_2^+$  &  $b_0^0$   & $b_1^0$ \\
                \hline
                \Vub &  \phantom{$-$}1.000 &&  &  & & \\
                $b_0^+$ & $-$0.791 &  \phantom{$-$}1.000 &&  &  & \\
                $b_1^+$&  \phantom{$-$}0.007  &$-$0.339 & \phantom{$-$}1.000 &&& \\
                $b_2^+$ &  \phantom{$-$}0.243 & $-$0.375 &$-$0.448 & \phantom{$-$}1.000 && \\
                $b_0^0$ & $-$0.376 &  \phantom{$-$}0.430 &$-$0.065 &$-$0.190 & \phantom{$-$}1.000 & \\
                $b_1^0$ &  \phantom{$-$}0.003 & $-$0.164 &  \phantom{$-$}0.127 &\phantom{$-$}0.244 &$-$0.830 & \phantom{$-$}1.000 \\
			\hline
			\hline
		\end{tabular}
		\label{tab:LCSR_pi}
	\end{center} 
\end{table*}

\begin{table*}[ht!]
	\begin{center}
	\caption{Full correlation matrix of \Vub and the BSZ form-factor coefficients from the fit to the \Btorholnu spectrum with LCSR constraints.}
		\begin{tabular}{cccccccc}
		    \hline
		    \hline
                &  \Vub  &   $b_0^{A_1}$ &  $b_1^{A_1}$   & $b_0^{A_2}$  &  $b_1^{A_2}$  & $b_0^{V}$ & $b_1^{V}$ \\
                \hline
                \Vub &  \phantom{$-$}1.000 &&  &  & & &\\
                $b_0^{A_1}$ & $-$0.464 &  \phantom{$-$}1.000 &&  & && \\
                $b_1^{A_1}$&  \phantom{$-$}0.035  & \phantom{$-$}0.542 & \phantom{$-$}1.000 &&&& \\
                $b_0^{A_2}$ &  $-$0.735 & \phantom{$-$}0.241 &$-$0.117 & \phantom{$-$}1.000 &&& \\
                $b_1^{A_2}$ & $-$0.126 &  $-$0.007 & \phantom{$-$}0.023 & \phantom{$-$}0.472 & \phantom{$-$}1.000 && \\
                $b_0^{V}$& $-$0.473 & \phantom{$-$}0.894 &  \phantom{$-$}0.493 &\phantom{$-$}0.255 &$-$0.056 & \phantom{$-$}1.000 \\
                $b_1^{V}$&  \phantom{$-$}0.064 & \phantom{$-$}0.538 &  \phantom{$-$}0.946 &$-$0.144 & \phantom{$-$}0.127 & \phantom{$-$}0.558 & \phantom{$-$}1.000 \\
			\hline
			\hline
		\end{tabular}
		\label{tab:LCSR_rho}
	\end{center} 
\end{table*}

\section{\Vub stability test}
\label{app:Vub}

\begin{figure*}[ht!]
    \begin{subfigure}[c]{0.325\textwidth}
		\centering
		\includegraphics[width=0.99\linewidth]{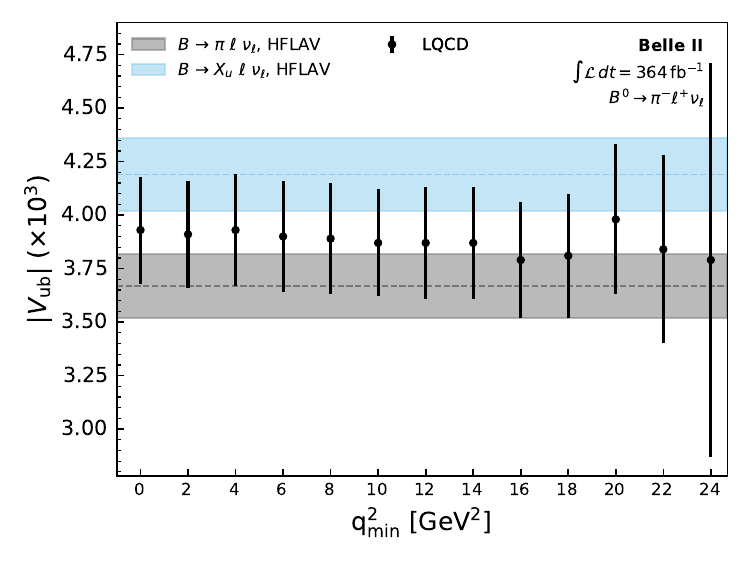}
		\subcaption{}
	\end{subfigure}
	\begin{subfigure}[c]{0.325\textwidth}
		\centering
		\includegraphics[width=0.99\linewidth]{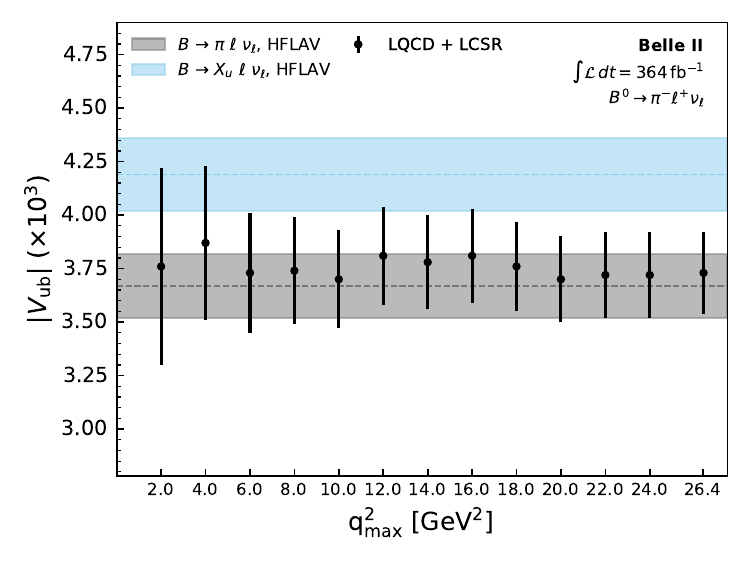}
		\subcaption{}
	\end{subfigure}
	\begin{subfigure}[c]{0.325\textwidth}
		\centering
		\includegraphics[width=0.99\linewidth]{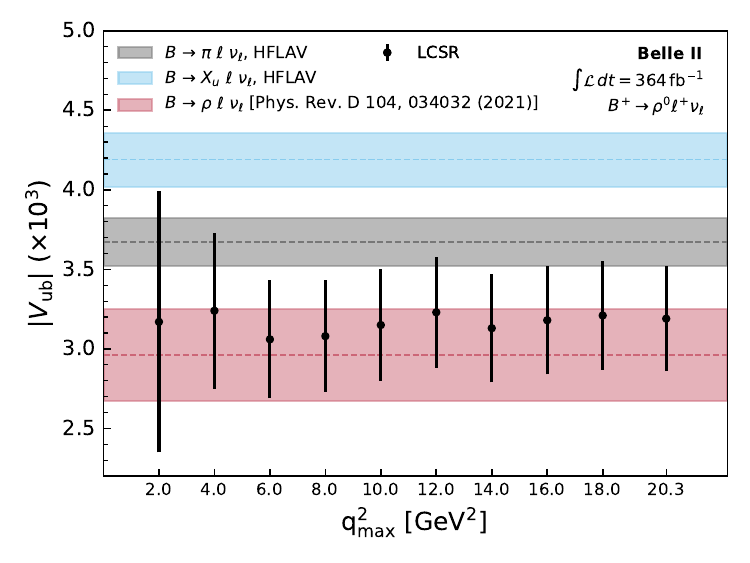}
		\subcaption{}
	\end{subfigure}
	\caption{Measured \Vub values from \Btopilnu{} (a,b) and \Btorholnu (c) for different cut-off values of \qtwo in the fit. In (a), where only LQCD constraints are used, minimum cut-off values are tested, while in (b,c), where LCSR constraints are used, maximum cut-off values are tested. 
    We also show a comparison to the world-average \Vub values from \totpilnu and inclusive \BtoXulnu from Ref.~\cite{HFLAV}. In (c) we further compare to the \Vub result obtained in Ref.~\cite{rho_om_FF} from \totrholnu measurements. }
    \label{fig:Vub_stab}
\end{figure*}

Fig.~\ref{fig:Vub_stab} presents the values of \Vub extracted when different \qtwo cut-off values are used during the $\chi^2$ fits.

\end{document}